\documentclass[journal,twoside]{IEEEtran}
\usepackage{color}
\usepackage{subfigure}
\usepackage{graphicx}
\usepackage{hyperref} 
\usepackage{booktabs}
\usepackage{longtable}
\usepackage{amssymb}
\usepackage{arabtex}
\usepackage{amsmath}
\usepackage{cite}
\usepackage{multirow}
\usepackage{url}
\usepackage{color}
\usepackage{bm}

\begin{document}

\title{Designs and Implementations in Neural Network-based Video Coding}


\author{Yue~Li,
        Junru~Li,
        Chaoyi~Lin,
        Kai~Zhang, \IEEEmembership{Senior Member, IEEE},
        and~Li~Zhang, \IEEEmembership{Senior Member, IEEE},
        Franck~Galpin,
        Thierry~Dumas,
        Hongtao~Wang,
        Muhammed~Coban,
        Jacob Str\"om,
        Du~Liu,
        Kenneth~Andersson
\thanks{Date of current version August 28, 2023. Thanks to the experts from JVET for their contributions to the development of NNVC. Thanks to the organizations providing test sequences for the verification of NNVC techniques.}
\thanks{Y. Li is with the Bytedance Inc., San Diego, CA 92122, USA (e-mail: yue.li@bytedance.com).}
\thanks{J. Li, C. Lin, K. Zhang, and L. Zhang are with Bytedance Inc. (e-mail: lijunru@bytedance.com; linchaoyi.cy@bytedance.com; zhangkai.video@bytedance.com; lizhang.idm@bytedance.com).}
\thanks{F. Galpin and T. Dumas are with InterDigital Inc. (e-mail: Franck.Galpin@interdigital.com; Thierry.Dumas@interdigital.com).}
\thanks{H. Wang and M. Coban are with Qualcomm Inc. (e-mail: hongtaow@qti.qualcomm.com; mcoban@qti.qualcomm.com).}
\thanks{J. Str\"om, D. Liu and K. Andersson are with Ericsson. (e-mail: jacob.strom@ericsson.com, du.liu@ericsson.com; kenneth.r.andersson@ericsson.com).}
}


\maketitle

\begin{abstract}
The past decade has witnessed the huge success of deep learning in well-known artificial intelligence applications such as face recognition, autonomous driving, and large language model like ChatGPT. Recently, the application of deep learning has been extended to a much wider range, with neural network-based video coding being one of them. Neural network-based video coding can be performed at two different levels: embedding neural network-based (NN-based) coding tools into a classical video compression framework or building the entire compression framework upon neural networks. This paper elaborates some of the recent exploration efforts of JVET (Joint Video Experts Team of ITU-T SG 16 WP 3 and ISO/IEC JTC 1/SC29) in the name of neural network-based video coding (NNVC), falling in the former category. Specifically, this paper discusses two major NN-based video coding technologies, i.e. neural network-based intra prediction and neural network-based in-loop filtering, which have been investigated for several meeting cycles in JVET and finally adopted into the reference software of NNVC.
Extensive experiments on top of the NNVC have been conducted to evaluate the effectiveness of the proposed techniques. 
Compared with VTM-11.0\_{nnvc}\footnote{VTM-11.0\_{nnvc} is the anchor for evaluating NNVC techniques, i.e. NNVC software with all NN-based tools off. It is equivalent to VTM-11.0 + enabled MCTF including the update from JVET-V0056 \cite{Wennersten2021Gop} + enabled deblocking in RDO \cite{Hu2019Encoder} + high level syntaxes for NNVC. NNVC software could be found in \url{https://vcgit.hhi.fraunhofer.de/jvet-ahg-nnvc/VVCSoftware_VTM}}, the proposed NN-based coding tools in NNVC-4.0 could achieve \{11.94\%, 21.86\%, 22.59\%\}, \{9.18\%, 19.76\%, 20.92\%\}, and \{10.63\%, 21.56\%, 23.02\%\} BD-rate reductions on average for \{Y, Cb, Cr\} under random-access, low-delay, and all-intra configurations respectively.
\end{abstract}


\IEEEpeerreviewmaketitle


\section{Introduction}
\label{sec:introduction}
With the popularization of smart phones and rapid development of video-based applications, the volume of video material has been increasing at a unprecedented speed in recent years. 
The efficient storage and transmission of mass data have become a great challenge.
To cope with this challenge, the Joint Video Experts Team of ITU-T SG 16 WP 3 and ISO/IEC JTC 1/SC29 has developed and finalized the latest video coding standard, namely Versatile Video Coding (VVC), to provide a more compact representation of video data \cite{bross2021overview}.  
VVC/H.266 has made significant progresses in terms of coding efficiency, providing approximately a 50\% bit-rate saving for equivalent perceptual quality relative to the performance of the prior standard High Efficiency Video Coding (HEVC)/H.265 \cite{sullivan2012overview}. While VVC offers a new level of capability for video compression, the necessity of developing more advanced video coding techniques still exists.

Classical video coding schemes epitomized by VVC adopt a sophisticated framework comprising numerous manually optimized and hand-crafted coding tools.
After development of several generations of video coding standards such as AVC/H.264 \cite{wiegand2003overview}, HEVC/H.265, and VVC/H.266, further improvement has become more and more difficult along this path. Therefore, experts are exploring other learning-based schemes to improve coding efficiency.

Due to the availability of powerful computing resources and abundant training data, deep learning has made a significant breakthrough in well-known artificial intelligence applications such as face recognition \cite{parkhi2015deep,wang2021deep}, autonomous driving \cite{huval2015empirical,rao2018deep}, and large language model like ChatGPT \cite{van2023chatgpt} in the past decade. Recently, the application of deep learning has been extended to a much wider range, especially to scenarios which can be easily formulated as a supervised problem. The target of video compression can be conceptualized as constructing a mapping from an original space (i.e., raw video data) to a latent domain (i.e., a bit stream), and back again, fitting the scope of deep learning. 
There exists two ways to build the mapping: utilizing both deep learning-based modules and non-learning-based modules, or utilizing purely deep learning-based modules \cite{liu2020deep,ma2019image}. Accordingly, the efforts exploring neural network-based video coding are distributed in two categories: embedding neural network-based (NN-based) coding tools into a classical video compression framework \cite{sun2020enhanced,dumas2020iterative,zhao2022efficient,yan2018convolutional,liu2019deep,yang2020deep,song2017neural,Zhou2018Convolutional,song2018practical,yang2023joint,li2017convolutional,kathariya2022multi}, or building the entire compression framework upon neural networks \cite{lu2019dvc,lin2020m,yang2020learning,li2021deep,sheng2022temporal,agustsson2020scale}.
In the former category, people usually design NN-based alternatives, e.g. NN-based intra/inter predictor \cite{sun2020enhanced, dumas2020iterative, zhao2022efficient, yan2018convolutional, liu2019deep}, transform \cite{yang2020deep}, arithmetic probability estimator \cite{song2017neural}, in-loop/post filter \cite{Zhou2018Convolutional, song2018practical, yang2023joint}, re-sampler \cite{li2017convolutional}, etc. to compete with the non-NN-based counterparts within the classical coding framework and rely on rate-distortion optimization to guarantee an improved performance.
While in the latter category, people adopt predictive coding-based method \cite{lu2019dvc, lin2020m, yang2020learning}, which first generates the predicted frame e.g. by using optical flow and then encodes residue e.g. with auto-encoder, or conditional coding-based method \cite{li2021deep, sheng2022temporal}, where prediction is embedded into the latent domain of auto-encoder.     

This paper elaborates some of the recent exploration efforts in JVET focusing on developing neural network-based video coding (NNVC) technologies beyond the capabilities of VVC. After investigation activities of several meeting cycles, the experts in JVET have identified two promising NN-based tools as an enhancement of conventional modules in the existing VVC design, i.e. neural network-based intra prediction and neural network-based in-loop filtering, and adopted them into the reference software of NNVC to demonstrate a reference implementation of encoding techniques, decoding process, as well as the training methods for these tools.
To generate a better intra prediction, a nonlinear mapping from causal neighboring samples to a prediction of the current block is derived using fully connected neural networks \cite{Dumas2023Neural,dumas2020iterative}. In addition, the neural network yields side outputs beneficial for subsequent Most Probable Mode (MPM) list construction and the transform kernel selection processes. To better recover details lost during compression, a convolutional neural network-based in-loop filter is designed \cite{Li2022Deepfixedpoint,Li2022rdo,Liu2023combined,Li2023additional,li2023idam}. The deep filter is trained iteratively to address the over-filtering issue \cite{li2023idam}. To further improve performance, the deep filter design also considers elements including coded information exploitation, parameter selection, inference granularity adaptation, residual scaling, temporal filtering, combination with deblocking filtering, harmonization with RDO, etc.

Extensive experiments on top of the Versatile Video Coding have been conducted to evaluate the techniques included in NNVC. 
Compared with VTM-11.0\_{nnvc}, the proposed NN-based coding tools in NNVC-4.0 could achieve \{11.94\%, 21.86\%, 22.59\%\}, \{9.18\%, 19.76\%, 20.92\%\}, and \{10.63\%, 21.56\%, 23.02\%\} BD-rate reductions on average for \{Y, Cb, Cr\} under random-access, low-delay, and all-intra configurations respectively.

The remainder of the paper is organized as follows. 
Section \ref{sec:intra_pred} introduces NN-based intra prediction technique. 
Section \ref{sec:NNLF} elaborates NN-based in-loop filtering technique. 
Section \ref{sec:sadl} describes the small ad-hoc deep learning (SADL) library for inference of NN-based models in NNVC.
Performance evaluation of NNVC techniques is presented in Section \ref{sec:experiment}.
Finally, Section \ref{sec:conclusion} concludes the paper.

\section{NN-based Intra Prediction}
\label{sec:intra_pred}

\begin{figure*}
\centering
\includegraphics[width=0.6\linewidth]{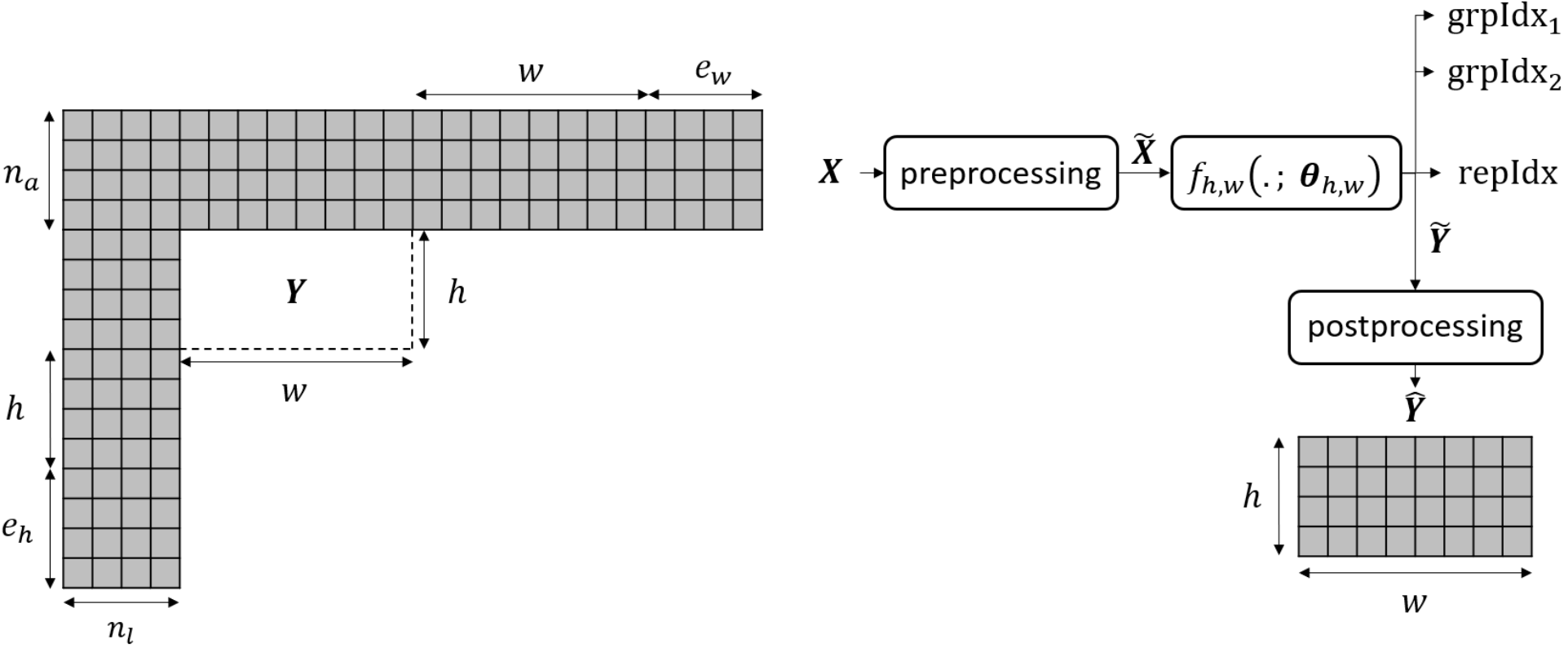}
\caption{Prediction of the current $w\times h$ block $\bm{Y}$ from the context $\bm{X}$ of decoded reference samples around $\bm{Y}$ via the neural network $f_{h, w} \left( \; . \; ; \; \bm{\theta}_{h, w} \right)$. In this figure, $h=4$, $w=8$, and $n_{a} = n_{l} = e_{h} = e_{w} = 4$.}
\label{fig_intra_framework}
\end{figure*}

\subsection{VVC coding tools directly interacting with the NN-based intra prediction}
\label{sec:coding_tools_in_vvc}
To justify the design of the NN-based intra prediction in Section \ref{sec:intra_framework}, the VVC coding tools featuring the strongest interactions in terms of compression efficiency with the NN-based intra prediction mode to be put into VVC must be detailed.

Any intra prediction mode, including a NN-based one, interacts in particular with the other intra prediction modes partly because the entropy coding in the signaling of the index of the intra prediction mode selected to predict a given block creates competition between them. Moreover, any intra prediction mode depends on transform coding as the residue resulting from the intra prediction of a given block is passed on to transform coding \cite{transform2021tcvst}. This dependency grows even more in the case of the secondary transforms in VVC, called Low-Frequency Non-Separable Transform (LFNST), as different LFNST kernels are specialized to different intra modes.

Precisely, for a given block predicted in intra and using the Discrete Cosine Transform-2 (DCT-2) horizontally and the DCT-2 vertically as primary transform, LFNST consists in applying a non-separable transform to the top-left region of the block of coefficients arising from the primary transform \cite{transform2021tcvst, Koo2019Low}. LFNST gathers $4$ transform sets with $2$ kernels per set. Note that, for a given kernel in a given transform set, the used matrix of weights and the shape of the top-left region involved in the second transform are determined by the size of the current block. The signaling of LFNST is decomposed into a so-called explicit signaling of the kernel set index and a so-called implicit signaling of the transform set index. In the explicit signaling, \textit{lfnstIdx} $\in \{ 0, 1, 2 \}$ is written to the bitstream. \textit{lfnstIdx} $= 0$ means that LFNST does not apply for the current block whereas, if \textit{lfnstIdx} $\{ 1, 2 \}$, \textit{lfnstIdx} - 1 indicates the used kernel set index. In the implicit signaling, the index of the intra prediction mode selected to predict the current block directly maps to the transform set index and whether the block of primary transform coefficients is transposed before applying LFNST. As no relationship between the \textit{directionality} of the prediction of a given block via a NN and the index of the NN-based intra prediction mode exists \cite{dumas2019Combined}, this mapping must not be reused for a block predicted via NN, and rather be produced by the NN, see Section \ref{sec:intra_framework}.

\subsection{Framework}
\label{sec:intra_framework}
The NN-based intra prediction mode contains $7$ neural networks, each predicting blocks of a different size in $\{ 4 \times 4, 8 \times 4, 16 \times 4, 32 \times 4, 8 \times 8, 16 \times 8, 16 \times 16 \}$.

In this NN-based intra prediction mode, the neural network predicting blocks of size $w\times h$ is denoted $f_{h, w} \left( \; . \; ; \; \bm{\theta}_{h, w} \right)$ where $\bm{\theta}_{h, w}$ gathers its parameters. For a given $w \times h$ block $\bm{Y}$ to be predicted, $f_{h, w} \left( \; . \; ; \; \bm{\theta}_{h, w} \right)$ takes a preprocessed version $\widetilde{\bm{X}}$ of the context $\bm{X}$ made of $n_{a}$ rows of $n_{l} + 2w + e_{w}$ decoded reference samples located above this block and $n_{l}$ columns of $2h + e_{h}$ decoded reference samples located on its left side to provide $\widetilde{\bm{Y}}$, see Fig.~\ref{fig_intra_framework}. The application of a postprocessing to $\widetilde{\bm{Y}}$ yields a prediction $\hat{\bm{Y}}$ of $\bm{Y}$.
The above-mentioned preprocessing and postprocessing are fully specified in Section \ref{sec:intra_pre_post}. Besides, to replace the mapping in the LFNST implicit signaling presented in Section \ref{sec:coding_tools_in_vvc}, $f_{h, w} \left( \; . \; ; \; \bm{\theta}_{h, w} \right)$ returns two indices $\text{grpIdx}_{1}$ and $\text{grpIdx}_{2}$. For $i \in \{1, 2\}$, $\text{grpIdx}_{i}$ denotes the index characterizing the LFNST transform set index and whether the primary transform coefficients resulting from the application of the DCT-2 horizontally and the DCT-2 vertically to the residue of the neural network prediction are transposed when $\text{lfnstIdx} = i$. Furthermore, for efficient synergy between the VVC intra prediction modes, i.e. PLANAR, DC, and the $65$ directional intra prediction modes, and the NN-based intra prediction, c.f. Section \ref{sec:MPM}, $f_{h, w} \left( \; . \; ; \; \bm{\theta}_{h, w} \right)$ returns the index repIdx $\in [0,66]$ of the VVC intra prediction mode whose prediction of $\bm{Y}$ from one row of decoded reference samples above $\bm{Y}$ and one column of decoded reference samples on its left side is the closest to $\hat{\bm{Y}}$.

Note that $n_{a}$, $n_{l}$, $e_{w}$, and $e_{h}$ together define the shape of the context $\bm{X}$ of $\bm{Y}$. $n_{a}$, $n_{l}$, $e_{w}$, and $e_{h}$ depend on $h$ and $w$, these dependencies being further explained in Section \ref{sec:context}.

\begin{figure}
\centering
\includegraphics[width=0.6\linewidth]{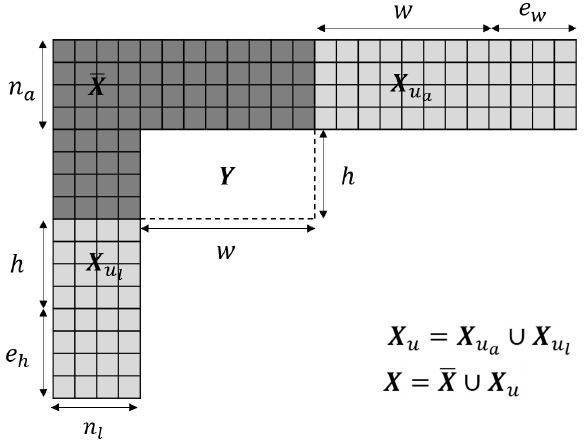}
\caption{Decomposition of the context $\bm{X}$ of decoded reference samples around the current $w\times h$ block $\bm{Y}$ into the available reference samples $\overline{\bm{X}}$ and the unavailable reference samples $\bm{X}_u$. In this figure, $h=4$, $w=8$, $n_{a} = n_{l} = e_{h} = e_{w} = 4$, and the number of unavailable reference samples reaches its maximum value.}
\label{fig_intra_preprocessing}
\end{figure}

\subsection{Preprocessing and Postprocessing}
\label{sec:intra_pre_post}
The preprocessing of the context fed into a neural network, shared by the training and test phases, is designed to obtain a range of values at the neural network input that eases optimization during the training phase \cite{Lecun1998efficient}. Precisely, the preprocessing in Fig.~\ref{fig_intra_framework} consists in the four following steps.
\begin{itemize}
 \item The mean $\mu$ of the available reference samples $\overline{\bm{X}}$ in $\bm{X}$ is subtracted from $\overline{\bm{X}}$, where the context $\bm{X}$ of the current $w\times h$ block $\bm{Y}$ is decomposed into the available reference samples $\overline{\bm{X}}$ and the unavailable reference samples $\bm{X}_u$, see Fig.~\ref{fig_intra_preprocessing}.
 \item The reference samples in the context $\bm{X}$ are multiplied by $\rho=1/(2^{b-8})$, $b$ being the internal bitdepth, i.e. 10 in VVC.
 \item All the unavailable reference samples $\bm{X}_u$ in $\bm{X}$ are set to 0.
 \item The context resulting from the previous step is flattened, yielding the vector $\widetilde{\bm{X}}$ of size $n_a(n_l+2w+e_w)+(2h+e_h)n_l$.
\end{itemize}

The postprocessing of the output of a neural network must approximatively reverses the above preprocessing. Precisely, the postprocessing depicted in Fig.~\ref{fig_intra_framework} consists in reshaping the vector $\widetilde{\bm{Y}}$ of size $hw$ into a rectangle of height $h$ and width $w$, dividing the result of the reshape by $\rho$, adding the mean $\mu$ of the available reference samples in the context of the current block, and clipping to $[0,\ 2^b-1]$. Therefore, the postprocessing can be summarized as
\begin{equation}
\label{eqn:intra_post}
\hat{\bm{Y}}=\min{\left(\max{\left(\frac{\text{reshape}{\left(\widetilde{\bm{Y}}\right)}}{\rho}+\mu,\ 0\right),\ 2^b-1}\right)}.
\end{equation}

Note that the above preprocessing and postprocessing apply to a neural network in floats. For a neural network in signed-integers exclusively, $\rho = 2^{Q_{in}-b+8}$, $Q_{in}$ denoting the input quantizer. For integer width $16$, $Q_{in} = 7$. For integer width $32$, $Q_{in} = 23$.

\subsection{MPM List Generation}
\label{sec:MPM}
In VVC, an efficient entropy coding of the index of the intra prediction mode selected to predict the current luma Coding Block (CB) involves a list of $6$ Most Probable Modes (MPMs). This list includes the index of the intra prediction mode selected to predict the luma CB above the current one and the index of the intra prediction mode selected to predict the luma CB on the left side of the current one.

In VVC with the NN-based intra prediction mode, if a non-NN-based intra prediction mode is selected to predict the current luma CB and the current luma CB is surrounded by luma CBs predicted via the NN-based intra prediction mode, the relevance of the list of MPMs of the current luma CB can be maintained thanks to repIdx. Indeed, if its left luma CB is predicted via the NN-based mode, the repIdx returned during the prediction of the left luma CB can become a candidate index to be put into the list of MPMs. If its above luma CB is predicted via the NN-based mode, the repIdx collected during the prediction of the above luma CB can become a candidate index to be put into the list of MPMs.

\subsection{Context Transformations}
\label{sec:context}
As said at the beginning of Section \ref{sec:intra_framework}, the NN-based intra prediction mode comprises the $7$ neural networks $\{f_{h, w} \left( \; . \; ; \bm{\theta}_{h, w} \right)\}_{\left( h, w \right) \in S}$, $S = \{ \left( 4, 4 \right), \left( 4, 8 \right), \left( 4, 16 \right)$, $\left( 4, 32 \right), \left( 8, 8 \right), \left( 8, 16 \right), \left( 16, 16 \right) \}$, each predicting blocks of corresponding shape in $S$. For a given $w\times h$ block to be predicted, the NN-based intra prediction mode may not contain $f_{h, w} \left( \; . \; ;\bm{\theta}_{h, w} \right)$. To circumvent this, context transformations help. Specifically, the context of the current block can be down-sampled vertically by a factor $\delta$ and/or down-sampled horizontally by a factor $\gamma$ and/or transposed before the step ``preprocessing'' in Fig.~\ref{fig_intra_framework}.
Then, the prediction of the current block can be transposed and/or up-sampled vertically by the factor $\delta$ and/or up-sampled horizontally by the factor $\gamma$ after the step ``postprocessing'' in Fig.~\ref{fig_intra_framework}. The transposition of the context of the current block and the prediction, $\delta$, and $\gamma$ are chosen so that a neural network belonging to the NN-based intra prediction mode can be picked for prediction, see Table \ref{tab:context_transformation}. Note that the NN-based intra prediction mode is disallowed for $(h, w)$ absent from Table \ref{tab:context_transformation}.

To limit the complexity of the neural network prediction, $n_{a} \left( h, w \right)$ and $n_{l} \left( h, w \right)$ are defined such that, after the potential context transformations, the number of rows and the number of columns in the resulting context never exceed $8$.

\begin{table}
\renewcommand{\arraystretch}{0.9}
\caption{Context transformations depending on the size of the block}
\label{tab:context_transformation}
\center
\begin{tabular}{l|c|c|c|c}
\hline
($h$, $w$)                                & $\gamma$     & $\delta$  & transposition  & neural network for prediction   \\
\hline
(4, 4)                                    & 1            & 1         & no             & $f_{4, 4}(.,\bm{\theta}_{4, 4})$     \\
\hline 
(4, 8)                                    & 1            & 1         & no             & $f_{4, 8}(.,\bm{\theta}_{4, 8})$     \\
\hline 
(8, 4)                                    & 1            & 1         & yes            & $f_{4, 8}(.,\bm{\theta}_{4, 8})$     \\
\hline 
(4, 16)                                   & 1            & 1         & no             & $f_{4, 16}(.,\bm{\theta}_{4, 16})$     \\
\hline 
(16, 4)                                   & 1            & 1         & yes            & $f_{4, 16}(.,\bm{\theta}_{4, 16})$     \\
\hline 
(4, 32)                                   & 1            & 1         & no             & $f_{4, 32}(.,\bm{\theta}_{4, 32})$     \\
\hline 
(32, 4)                                   & 1            & 1         & yes            & $f_{4, 32}(.,\bm{\theta}_{4, 32})$     \\
\hline 
(8, 8)                                    & 1            & 1         & no             & $f_{8, 8}(.,\bm{\theta}_{8, 8})$     \\
\hline 
(8, 16)                                   & 1            & 1         & no             & $f_{8, 16}(.,\bm{\theta}_{8, 16})$     \\
\hline                                        
(16, 8)                                   & 1            & 1         & yes            & $f_{8, 16}(.,\bm{\theta}_{8, 16})$     \\
\hline 
(8, 32)                                   & 2            & 1         & no             & $f_{8, 16}(.,\bm{\theta}_{8, 16})$     \\
\hline 
(32, 8)                                   & 1            & 2         & yes            & $f_{8, 16}(.,\bm{\theta}_{8, 16})$     \\
\hline 
(16, 16)                                  & 1            & 1         & no             & $f_{16, 16}(.,\bm{\theta}_{16, 16})$     \\
\hline 
(16, 32)                                  & 2            & 1         & no             & $f_{16, 16}(.,\bm{\theta}_{16, 16})$     \\
\hline 
(32, 16)                                  & 1            & 2         & no             & $f_{16, 16}(.,\bm{\theta}_{16, 16})$     \\
\hline 
(32, 32)                                  & 2            & 2         & no             & $f_{16, 16}(.,\bm{\theta}_{16, 16})$     \\
\hline 
(64, 64)                                  & 4            & 4         & no             & $f_{16, 16}(.,\bm{\theta}_{16, 16})$     \\
\hline 
\end{tabular}
\end{table}

\subsection{Signaling of the NN-based Intra Prediction Mode}
\label{sec:signalingintra}

\subsubsection{Signaling in luma}
\label{sec:signalingluma}
\begin{figure}
\centering
\includegraphics[width=0.6\linewidth]{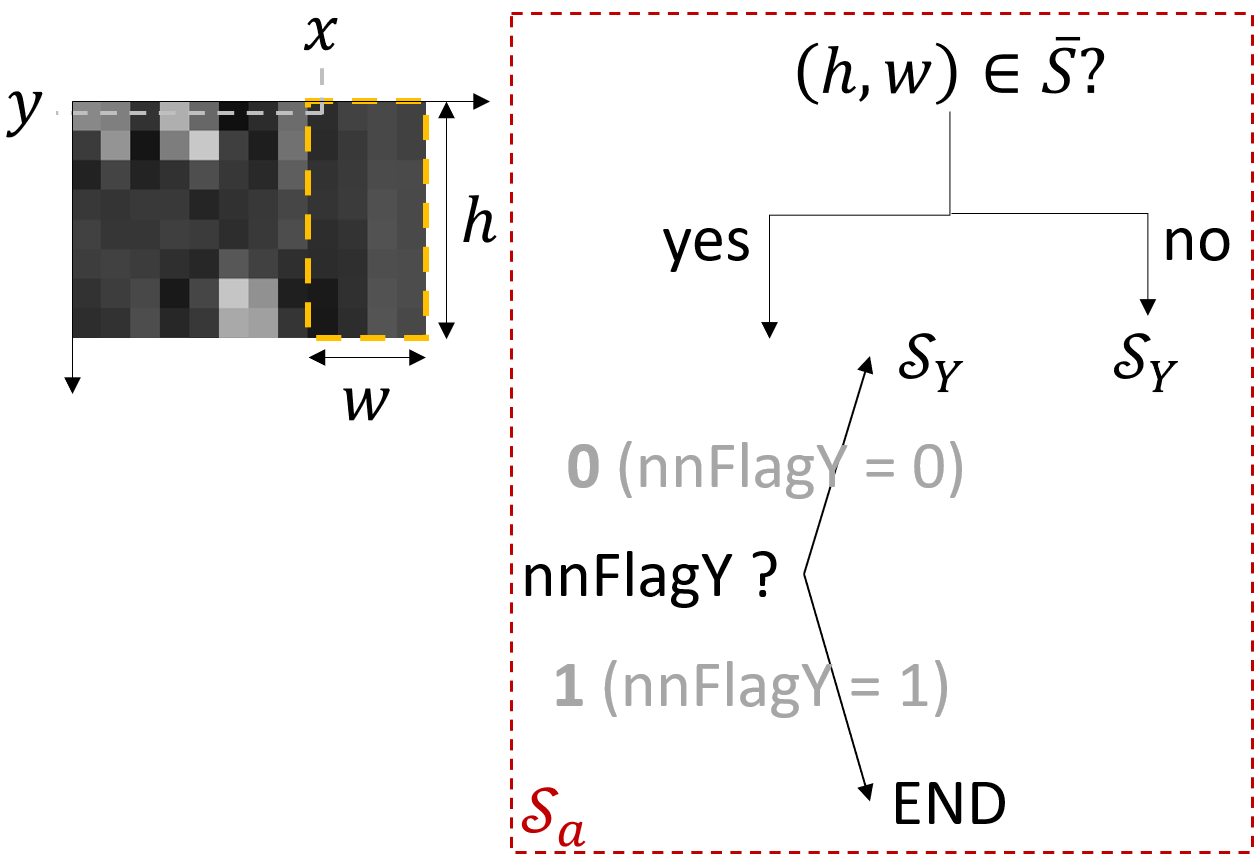}
\caption{Adapted intra prediction mode signaling $\mathcal{S}_{a}$ of the current $w \times h$ luma CB. This CB is framed in orange using dashed line. The bin value of $nnFlagY$ appears in bold gray.}
\label{fig_signaling_luma}
\end{figure}

Given that the NN-based intra prediction mode predicts blocks of each shape in $\overline{S} = S \; \cup \; \{ \left( 8, 4 \right), \left( 16, 4 \right)$, $\left( 32, 4 \right)$, $\left( 16, 8 \right)$, $\left( 8, 32 \right)$, $\left( 32, 8 \right)$, $\left( 16, 32 \right)$, $\left( 32, 16 \right)$, $\left( 32, 32 \right)$, $\left( 64, 64 \right) \}$, c.f. Section \ref{sec:context}, the intra prediction mode signaling of the current $w \times h$ luma CB can be adapted to incorporate the NN-based mode, at low cost, by introducing a flag $nnFlagY$ only if $\left( h, w \right) \in \overline{S}$. In details, the adapted intra prediction mode signaling $\mathcal{S}_{a}$ of the current $w \times h$ luma CB whose top-left pixel is at position $\left( y , x \right)$ in the current luma channel is split into two cases, see Fig. \ref{fig_signaling_luma}.
\begin{itemize}
\item If $\left( h, w \right) \in \overline{S}$, $nnFlagY$ appears. $nnFlagY = 1$ means that the NN-based mode is selected, then END. $nnFlagY = 0$ tells that the NN-based mode is not selected, then the regular VVC intra prediction mode signaling $\mathcal{S}_{Y}$ of the current luma CB applies.
\item Otherwise, $\mathcal{S}_{Y}$ applies.
\end{itemize}
Note that, in the case $\left( h, w \right) \in \overline{S} \; \&\& \; nnFlagY  = 1$, if the context of the current luma CB goes out of the current luma channel bounds, i.e. $x < n_{l} \; \| \; y < n_{a}$, PLANAR replaces the NN-based intra prediction.

\subsubsection{Signaling in chroma}
Before presenting the signaling of the NN-based mode in chroma, the Direct Mode (DM) in VVC must be detailed. For a given pair of chroma CBs predicted via the DM, the intra prediction mode selected to predict the luma CB being collocated with this pair of chroma CBs is used to predict each of these two chroma CBs \cite{intra2021tcvst}.

Based on the principle of the proposed signaling in luma, c.f. \ref{sec:signalingluma}, the adapted intra prediction mode signaling of the current pair of $w \times h$ chroma CBs whose top-left pixel is at position $\left( y , x \right)$ in the current pair of chroma channels is decomposed into two cases.
\begin{itemize}
    \item If the luma CB collocated with this pair of chroma CBs is predicted by the NN-based mode
    \begin{itemize}
        \item If $\left( h, w \right) \in \overline{S}$, denoted Case $\left[ * \right]$, the DM becomes the NN-based intra prediction mode.
        \item Otherwise, the DM is set to PLANAR.
    \end{itemize}
    \item Otherwise
    \begin{itemize}
        \item If $\left( h, w \right) \in \overline{S}$, $nnFlagC$ is placed before the DM flag in the decision tree of the intra prediction mode signaling in chroma. $nnFlagC = 1$, a.k.a Case $\left[ ** \right]$, indicates that the NN-based mode is selected, then END. $nnFlagC = 0$ tells that the NN-based mode is not selected, then the regular VVC intra prediction mode signaling $\mathcal{S}_{C}$ of the current pair of chroma CBs resumes from the DM flag.
        \item Otherwise, $\mathcal{S}_{C}$ applies.
    \end{itemize}
\end{itemize}
Note that, in Cases $\left[ * \right]$ and $\left[ ** \right]$, if the context of the current chroma CB goes out of the current chroma channel bounds, i.e. $x < n_{l} \; \| \; y < n_{a}$, PLANAR replaces the NN-based intra prediction mode.

\subsection{Training}
\begin{figure*}[th]
\centering
\includegraphics[width=0.75\linewidth]{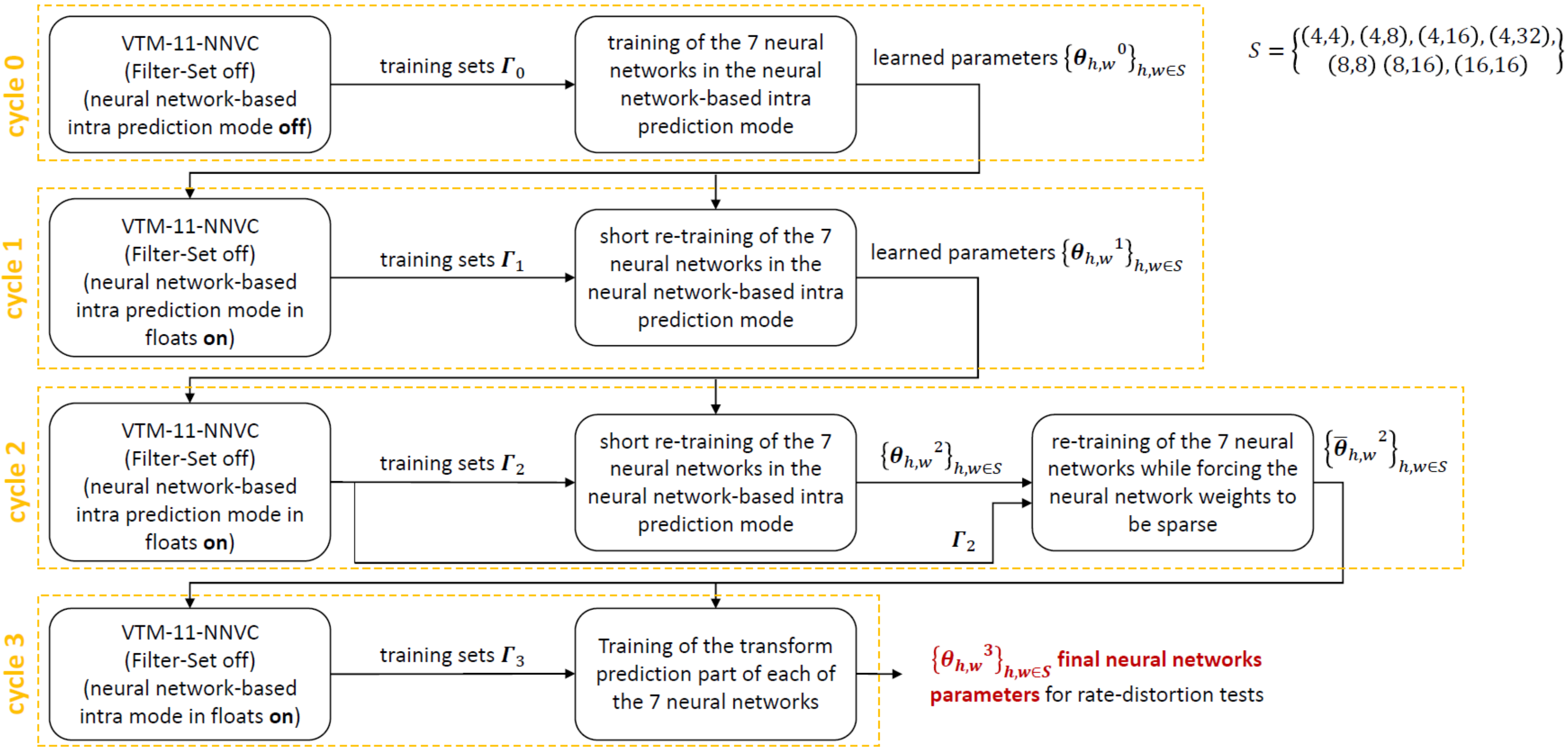}
\caption{Iterative training of the neural networks belonging to the NN-based intra prediction mode.}
\label{fig_it_intra}
\end{figure*}

In an Intra-slice (I-slice) in VVC, as the partitioning is mainly driven by the intra prediction, the reference-samples-to-block relationships specific to the VVC intra prediction modes are usually retrieved in the pairs of a partitioned block and its reference samples \cite{dumas2020iterative}. Thus, the training of a neural network on pairs a block extracted from the VVC partitioning of a given frame and its context leads to the neural network learning essentially the VVC intra prediction capability. 
To bypass this, an iterative training of neural networks for intra prediction is developed, see Fig. \ref{fig_it_intra}.

\begin{itemize}
\item At cycle $0$, VTM-11.0 anchor produces pairs of a block and its context. Then, the $7$ neural networks are trained on them, initializing their parameters randomly.
\item At cycle $1$, VTM-11.0 with the NN-based intra prediction mode using the parameters trained at cycle $0$ produces pairs of a block and its context. Then, the $7$ neural networks are trained on them, initializing their parameters from their state at the end of cycle $0$.
\item At cycle $2$, VTM-11.0 including the NN-based intra prediction mode using the parameters trained at cycle $1$ generates pairs of a block and its context. Then, the $7$ neural networks are trained on them, initializing their parameters from their state at the end of cycle $1$. Then, using the same training data, the trainings of these $7$ neural networks are resumed, introducing this time a sparsity constraint on their weights.
\item At cycle $3$, VTM-11.0 with the NN-based intra prediction mode using the parameters trained at cycle $2$ gives training data. Then, the portion computing $\text{grpIdx}_{1}$ and $\text{grpIdx}_{2}$ in each of the $7$ neural networks is trained on them, initializing their parameters from their state at the end of cycle $2$.
\end{itemize}

\subsection{Inference Details}
Small Ad-hoc Deep Learning library (SADL), c.f. Section \ref{sec:sadl}, runs the inference of the NN-based intra prediction, see Table \ref{tab:intra_inference}, using its fixed point-based implementation where both neurons and weights are represented as 16-bit signed integer. In each neural network, each intermediate representation features $1216$ neurons. In each layer, LeakyReLU is chosen as non-linearity, excluding the last layer without non-linearity.

\subsection{Relation to the state-of-the-art}
Prior to the proposed NN-based intra prediction mode in VVC, neural networks for intra prediction have been integrated into either VVC or one of its predecessor. Especially, in \cite{neuralNetworkBased2018SPIE, intraPicturePredictionDCC2019}, multiple neural networks are jointly trained and then integrated as a single intra prediction mode into a predecessor of VVC: HEVC enhanced with non-square partitions and VTM-1.0 respectively. Precisely, in this mode, a different set of neural network predicts blocks of each size. During the training phase, the use of a partitioner and an objective function being the minimum rate-distortion cost computed from the neural network predictions over all partitions of each training block into sub-blocks induces a specialization of different neural networks to different classes of textures. Note that the iterative simplifications of this NN-based intra prediction mode \cite{affineLinearWeightedIntra, simplificationsOfMip} has led to Matrix-based Intra Prediction (MIP), being part of VVC.

Note that, as MIP is a VVC intra prediction mode, the experiments $M_{1}$ in Table \ref{table_ablation_results} reflects the rate-distortion performance of the proposed NN-based intra prediction mode on top of MIP.

\begin{table}[h]
\renewcommand{\arraystretch}{0.9}
\caption{Inference information on the NN-based intra predictors}
\label{tab:intra_inference}
\center
\begin{tabular}{l|l}
\hline
Hardware type                             & single thread CPU            \\
\hline
Framework                                 & SADL                         \\
\hline
Parameter number                          & 1.52M in total               \\
\hline
Parameter precision (bits)                & 16                           \\
\hline
Worst-case kMAC/pixel                     & 7.7                           \\
\hline
Total convolutional layers               & 0                              \\
\hline
Total fully-connected layers              & 4 for (16, 16), 3 for others  \\
\hline
Batch size                                & 1                              \\
\hline
Patch size                                & see Section \ref{sec:context}  \\ 
\hline                                        
\end{tabular}
\end{table}

\section{NN-based In-loop Filter}
\label{sec:NNLF}
The proposed NN-based in-loop filter is known as filter set \#1 \cite{Li2022Deepfixedpoint} in NNVC-4.0. The filter architectures are introduced first, followed by an elaboration on parameter selection, residual scaling, temporal filtering, harmonization with RDO, etc. At last, we describe the inference and training details of the filter.

\begin{figure*}
\centering
\includegraphics[width=0.65\linewidth]{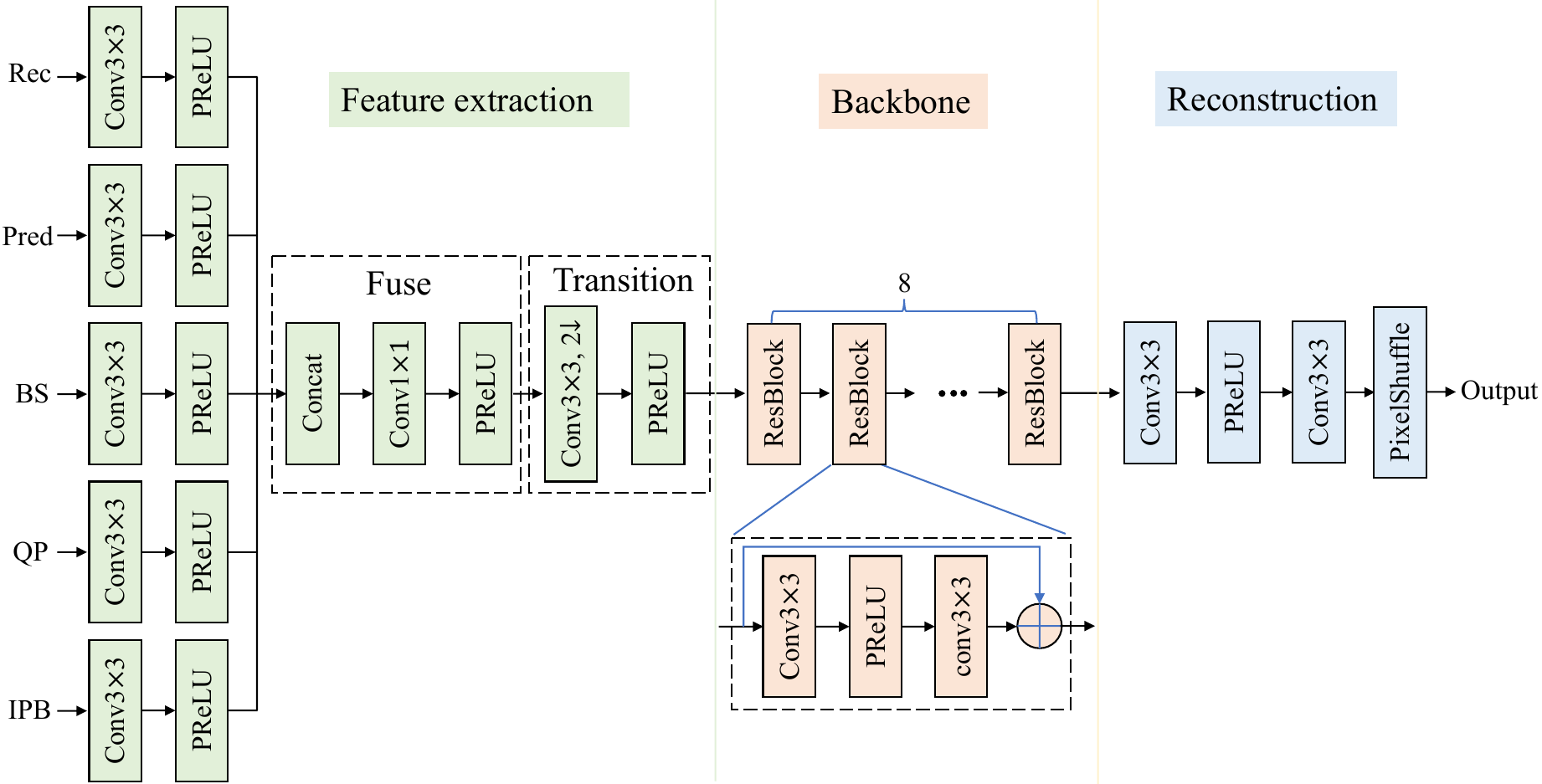}
\caption{Schematic of CNN-based in-loop filter. Rec, Pred, BS, QP, and IPB stand for reconstruction samples, prediction samples, boundary strength, quantization parameter and prediction types respectively. Number of feature maps is 96 for all internal layers.}
\label{fig_CNNLF}
\end{figure*}

\subsection{Network Architecture}
\label{sec:nnlf_arch}
Fig. \ref{fig_CNNLF} illustrates the diagram of CNN-based in-loop filter for luma component, comprising feature extraction, backbone, and reconstruction parts.

It is asserted in \cite{li2021convolutional} that applying existing in-loop filters in VVC prior to the CNN filter may cause the loss of important information.
Therefore, the reconstruction samples (Rec in Fig. \ref{fig_CNNLF}) refer to samples unfiltered by existing in-loop filters in VTM.
Note that in-loop filters such as SAO (sample adaptive offset, \cite{fu2012sample}) or ALF (adaptive loop filter, \cite{karczewicz2016geometry}) create nontrivial bitrate overhead to lower the compression distortion. When placed after the deep filter, this overhead may be reduced.
Besides reconstruction, auxiliary inputs are utilized to improve the performance. 
Intra/inter prediction is the key process for reducing spatial and temporal redundancy. The encoder selects a prediction mode with the best rate-distortion trade-off from a list of candidates during the encoding process. In other words, the prediction samples from the decoder side could reflect decisions made by the encoder, providing important clues about original samples. In addition, compression distortion is directly caused by the quantization on the residues in the transform domain, while residues are highly dependent on the quality of prediction samples, therefore prediction samples can also significantly impact the type and strength of artifacts in the decoded images. Taking the above analysis into account, prediction samples (Pred in Fig. \ref{fig_CNNLF}) are fed into the filter as an additional auxiliary input. Similarly, boundary strength (BS in Fig. \ref{fig_CNNLF}) generated during deblocking process reflects the strength of compression artifact near block boundaries. Inputs of QP and IPB make the filter aware of high-level compression conditions, i.e., quantization parameter and prediction types (intra, uni-inter, bi-inter). Given these two auxiliary inputs, a single model is capable of handling contents compressed with different QPs and block prediction types. 
For instance, if the IPB information states that a block has been inter predicted rather than intra predicted, this means that the block has likely been NN-filtered once already, and the NN model can lower the filtering strength to avoid over-filtering.

The feature extraction part accounts for aggregating informations from different inputs.
Specifically, individual features are extracted separately, concatenated together along the channel dimension, shrunk through a $1\times1$ convolutional layer, and downsampled to half resolution, to form a compact representation of all inputs. 
The network backbone, which is consisting of 8 cascaded residual blocks, transforms the compact representation into clean features with less compression artifacts.
At last, the reconstruction part maps the clean features into the pixel domain to predict the details lost during compression.
Note that the CNN-based in-loop filter is actually designed to learn the mapping from distorted input to lost details (residual between groundtruth and distorted input), thus the final output can be obtained by,
\begin{equation}
\label{eqn:output}
\bm{R}_{nn}=\bm{R}_{no} + f(\bm{R}_{no})
\end{equation}
where $\bm{R}_{no}$ and $\bm{R}_{nn}$ denote the unfiltered samples and filtered samples respectively, while $f$ is the CNN-based in-loop filter.

Coding tools such as CCLM \cite{zhang2018enhanced} and CCALF \cite{misra2019cross} utilize luma information for boosting chroma performance.
Similarly, the luma information is exploited for the chroma in-loop filtering. 
In the YUV 4:2:0 format where the luma resolution is higher than that for chroma, features are first extracted separately from luma and chroma. Then luma features are downsampled and concatenated with chroma features. 
Note that IPB information is not included in the chroma filter, as no benefits are observed from this input for chroma.
The same network backbone and reconstruction parts from the luma network are used for the chroma network.

\subsection{Parameter Selection}
\label{sec:param_select}
As analyzed in \cite{li2023idam}, the content propagation phenomenon that exists in the inter coding case may deteriorate the efficiency of in-loop filtering significantly, as samples filtered in one frame may be propagated to a following frame and filtered again, leading to over-filtering.
Intuitively, providing options with multiple filtering strengths may mitigate the over-filtering issue. 
To adjust the filtering strengths of a candidate filter, one possible way is to modify its input parameter QP slightly, because distortion levels at different QPs should cause filtering behaviors with corresponding strengths during the training process.


Without loss of generality, a candidate list containing three QP parameters is considered by default. Using more parameters may bring better performances at the cost of higher encoding complexity, and vice versa.
At encoder side, each picture or block could determine whether to apply the CNN-based in-loop filter or not. When the CNN-based filter is determined to be applied to a picture or a block, the QP parameter must be selected from a candidate list. Specifically, all blocks in the current picture are filtered using three QP parameters in the encoder.
Then five costs, i.e. \textit{Cost\_0}, ..., \textit{Cost\_5}, are calculated and compared against each other to find the best rate-distortion trade-off.
In \textit{Cost\_0}, CNN-based filter is prohibited for all blocks.
In \textit{Cost\_i}, \textit{i = 1, 2, 3}, CNN-based filter with \textit{$i^{th}$} parameter is used for all blocks.
In \textit{Cost\_4}, different blocks may prefer different parameters, and the information regarding whether to use CNN-based filter, and if so, which parameter to use is signaled for each block.
At decoder side, whether to use CNN-based filter and which parameter to use for a block is based on the \textit{Param\_Id} parsed from the bit-stream 

Denote the sequence level QP as q, the candidate list \{Param\_1, Param\_2, Param\_3\} is set as \{q, q-5, q-10\} and \{q, q-5, q+5\} for low and high temporal layers respectively. Stronger filtering strength is used for high temporal layers, because coarser quantization used in these layers may yield large distortion, and since higher temporal layers are used less for prediction, over-filtering is less of a problem. A shared parameter is used for the two chroma components to lower the worst-case complexity at the decoder side. In addition, the number of parameter candidates could be specified at the encoder side. For the all-intra configuration, the parameter selection is disabled while filter on/off control is still preserved, since there is no content propagation issue in this configuration.

For further improving the adaptation capability, granularity (block size) of the on/off control and the parameter selection is made dependent on resolution and bitrate.
For a higher resolution, the granularity will be coarser as content tends to change slower, and vice versa.
For a higher bitrate, the granularity will be finer since more overhead bits can be afforded, and vice versa.

\subsection{Residual Scaling}
\label{sec:RS}
As pointed out in Section \ref{sec:param_select}, varying the filtering strength according to picture content may alleviate the over-filtering issue.
Residual scaling is another mechanism (besides parameter selection) to achieve the purpose of filtering strength adjustment, and can be formulated as,
\begin{equation}
\label{eqn:RS}
\bm{R}_{nn}=\omega \cdot (\bm{R}_{nn} - \bm{R}_{db}) + \bm{R}_{db}
\end{equation}
where $\bm{R}_{db}$ is the deblocking filtered samples, $\omega$ is the scaling factor derived based on least square method.
(\ref{eqn:RS}) indicates that the residual between deblocking filtered samples and NN filtered samples can be scaled by a scaling factor and then added back to the deblocking filtered samples. For each color component, a scaling factor is signaled.
It is worth noting that (\ref{eqn:RS}) can be written as,
\begin{equation}
\label{eqn:combination}
\bm{R}_{nn}=\omega \cdot \bm{R}_{nn} + (1 - \omega) \cdot \bm{R}_{db}
\end{equation}
(\ref {eqn:combination}) implies a convex combination of NN filtering and deblocking filtering. It is asserted that using $\bm{R}_{db}$ instead of $\bm{R}_{no}$ in (\ref {eqn:combination}) benefits perceptual quality \cite{Andersson2022ComDbNn}. 
The reason is that the NN filter has the effect of removing deblocking artifacts, but if turned off in (4) (i.e., $\omega=0$), the result would be an output without any deblocking, given $\bm{R}_{db}$ was replaced by $\bm{R}_{no}$. In contrast, (\ref {eqn:combination}) guarantees that the output will be deblocked one way or the other, either through the NN filter or through the regular deblocking filter.

\subsection{Temporal Filtering}
\label{sec:temporal_filtering}
In video coding, neighboring reconstructed pictures might have a higher quality than the current picture since quality fluctuation usually exists across compressed pictures. 
This has motivated work on multi-frame quality enhancement \cite{guan2019mfqe}, which take advantage of adjacent pictures with higher quality to enhance the current picture.

\begin{figure}
\centering
\includegraphics[width=0.7\linewidth]{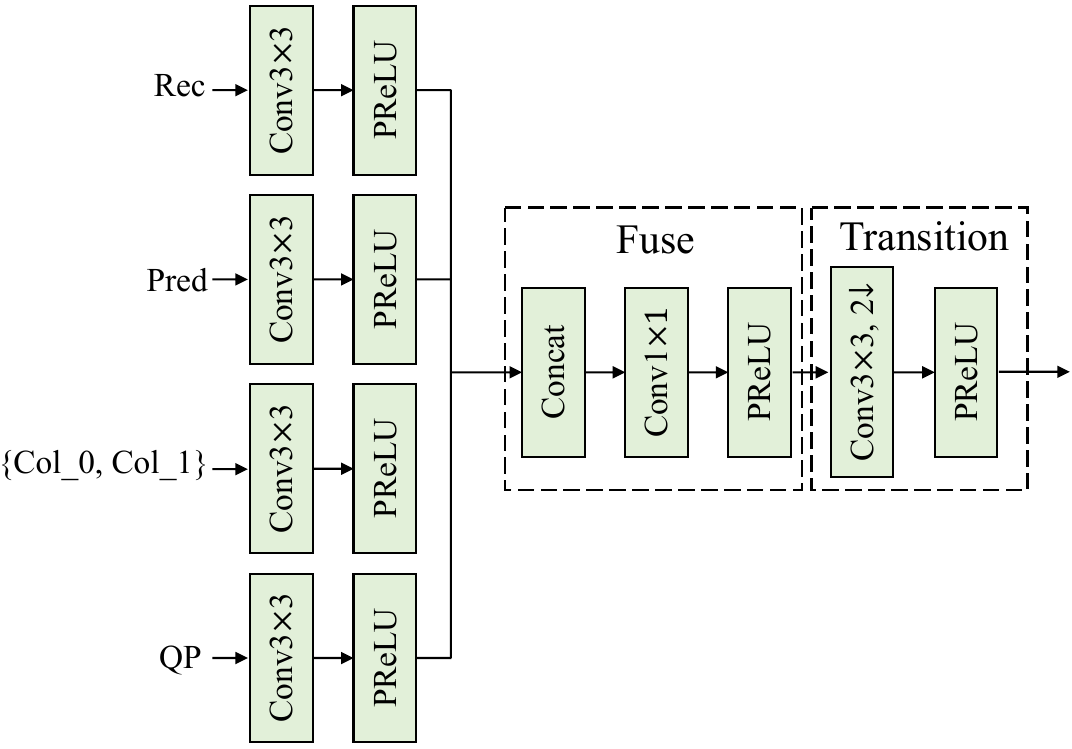}
\caption{Temporal in-loop filter. Only feature extraction part is illustrated, other parts remain the same as in Fig.~\ref{fig_CNNLF}. \{Col\_0, Col\_1\} refers to collocated samples from the first picture in both reference picture lists.}
\label{fig_temporal}
\end{figure}

Since the hierarchical coding structure under the random-access configuration naturally leads to previously coded pictures of higher quality, the temporal information from the reference picture can be exploited for the in-loop filtering of current picture. 
Specifically, NNVC includes an additional in-loop filter as shown in Fig.~\ref{fig_temporal}, namely temporal filter, taking collocated blocks from the first picture in both reference picture lists to improve performance.
Note that, to avoid complexity, the two collocated blocks are directly concatenated and fed into the temporal filter without any explicit temporal alignment operations.
In addition, this temporal filter is only activated for pictures in the three highest temporal layers, because in low temporal layers, the temporal correlation between collocated blocks and the current block is weak and thus limits the performance. 

\subsection{Encoder Optimization with NN Filter}
\label{sec:EncNnlfOpt}
Rate-distortion optimization (RDO) of the partitioning tree structure plays a vital role to to increase coding efficiency, both for traditional codecs such as VVC as well as for NNVC.
However, in NNVC, there exists a gap between the reconstruction samples used for distortion calculation during the RDO and the ultimate reconstruction samples, since the latter are eventually filtered using the NN model.
To bridge this gap, a NN filter is inserted into the RDO process for partitioning mode selection. 

Specifically, a NN filter is applied on the reconstruction samples before comparing them with the original samples to calculate the distortion. The optimal partitioning mode is then selected based on the refined RD cost. 
To reduce complexity, several fast algorithms are introduced. 
First, instead of using the full NN in-loop filter, an aggressively simplified version of the NN filter is used. 
Second, the parameter selection is omitted. 
Third, coding unit allowing using this technique should have a size no larger than 64. 
At last, the refined cost will be used only if the difference to the original R-D costs lies in a predefined range.

\subsection{Training Details}
\begin{figure*}
\centering
\includegraphics[width=0.6\linewidth]{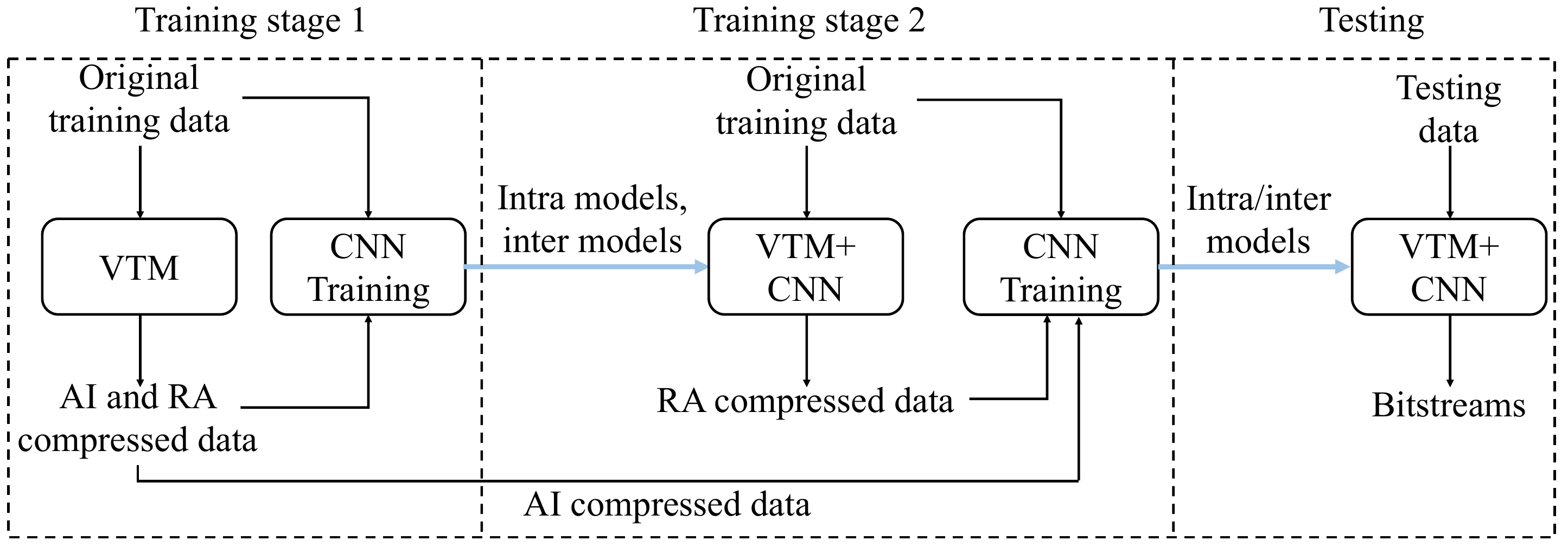}
\caption{Iterative training of CNN-based in-loop filter with two stages.}
\label{fig_it}
\end{figure*}

An iteratively conducted two-stage method is adopted to train the NN-based in-loop filters as shown in Fig.~\ref{fig_it}. 
The iterative training explicitly takes the filtering effect on reference frames into account during training process, in order to ease the over-filtering issue described in Section \ref{sec:param_select}.

\begin{itemize}
\item In the first stage, NNVC with NN filtering disabled (equivalent to the VTM anchor) is used to compress training images and videos under all-intra and random-access configurations. The reconstructed images and videos together with other auxiliary information are collected and utilized for training intra and inter filters.

\item In the second stage, NNVC equipped with the models from the previous training stage is used to compress the training videos in the random-access setting. That is to say, intra pictures and inter pictures will be processed by the intra filters and inter filters obtained in training stage 1, respectively. Then, the intra training data from stage 1 and inter training data from stage 2 are combined to train the unified intra and inter models (one model for both intra luma and inter luma, one model for both intra chroma and inter chroma).
\end{itemize}

Note that the temporal filter can be trained using a similar strategy. Training images and videos are from the DIV2K dataset \cite{timofte2017ntire} and the BVI-DVC dataset \cite{ma2021bvi}. PyTorch \cite{paszke2019pytorch} serves as the training platform. More details regarding training can be found in the NNVC training folder\footnote{\url{https://vcgit.hhi.fraunhofer.de/jvet-ahg-nnvc/VVCSoftware_VTM/-/tree/VTM-11.0_nnvc/training}}.

\subsection{Inference Details}
SADL (see Section \ref{sec:sadl}) is used for the inference of the NN-based in-loop filters. Both floating point-based and fixed point-based implementations are supported, however a real codec would need to use fixed-point arithmetics to avoid drift. In the fixed-point implementation, both weights and feature maps are represented with int16 precision using a static quantization method. In total, there are three filter models, i.e luma filter, chroma filter, and temporal filter. Other details regarding the inference is provided in Table \ref{tab:nnlf_inference}. 

\begin{figure}
\centering
\includegraphics[width=0.9\linewidth]{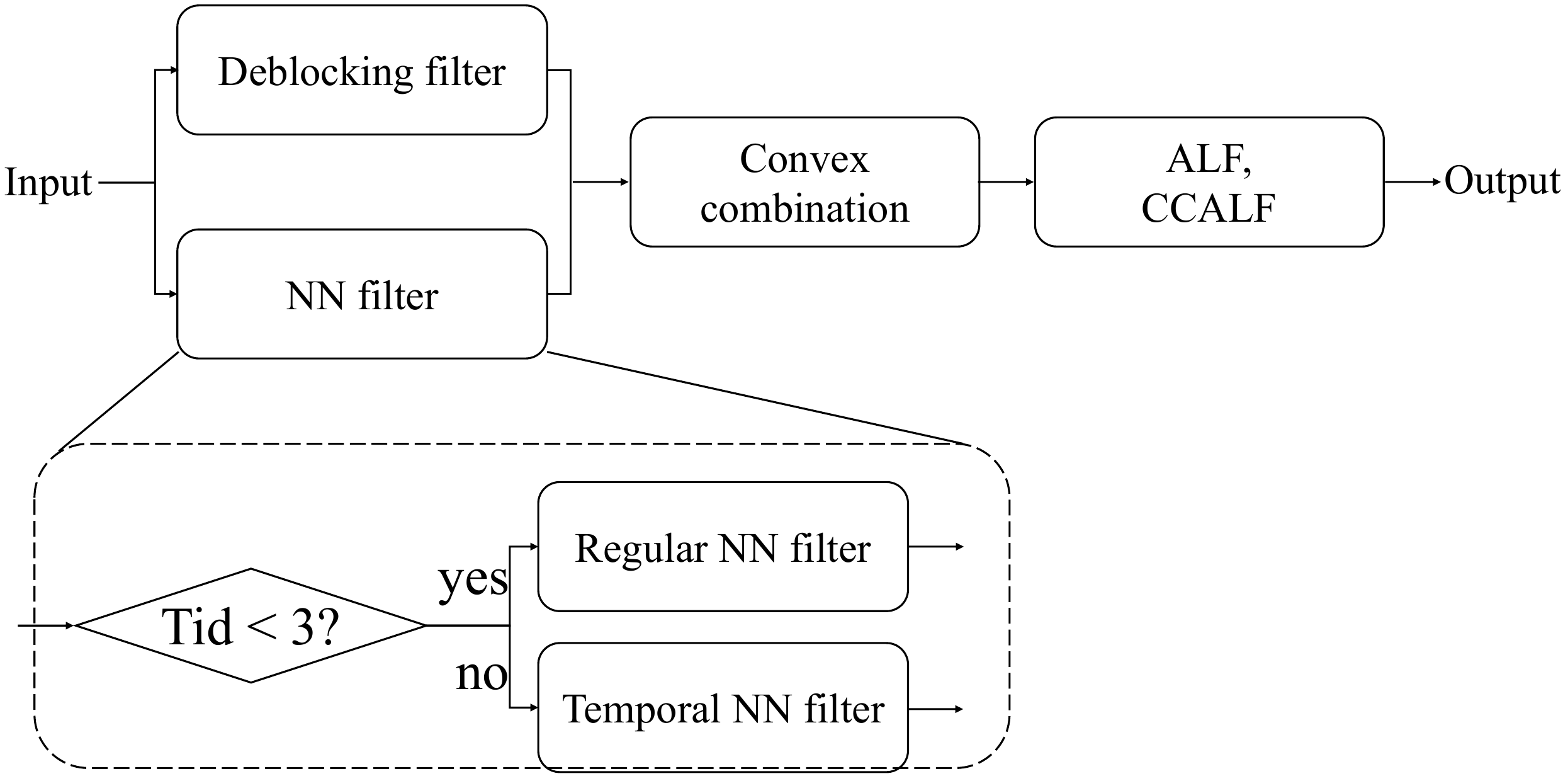}
\caption{Embedding of CNN-based in-loop filter into codec.}
\label{fig_embedding}
\end{figure}

Fig.~\ref{fig_embedding} depicts how to harmonize the NN-based filter with existing loop filters in VVC \cite{karczewicz2021vvc}. Deblocking and NN filtering are performed in parallel and then convexly combined via (\ref {eqn:combination}). SAO is disabled as no additional benefits are observed on top, while ALF and CCALF are placed after the CNN-based filtering to reduce overhead. As analyzed in Section \ref{sec:temporal_filtering}, temporal NN filter is proposed for pictures at high temporal layers ($Tid \geq 3$) while regular NN filter handles the others.

\begin{table}
\renewcommand{\arraystretch}{0.9}
\caption{Inference information of NN-based in-loop filters}
\label{tab:nnlf_inference}
\center
\begin{tabular}{l|l}
\hline
Hardware type                             & single thread CPU            \\
\hline
Framework:                                & SADL                                         \\
\hline
Parameter Number                          & 1.55M/model \\
\hline
Parameter Precision (Bits)                & 16                                      \\
\hline
Worst-case kMAC/pixel                     & 673             \\
\hline
Total Conv. Layers                        & 25                                           \\
\hline
Total FC Layers                           & 0                                            \\
\hline
Batch size:                               & 1                                            \\
\hline
Patch size                                & 128+16, 256+16                                          \\ 
\hline                                        
\end{tabular}
\end{table}

\section{Small Ad-hoc Deep-Learning Library}
\label{sec:sadl}
\subsection{Overview}
Small Ad-hoc Deep-Learning library (SADL) is a header-only small library for neural network inference available at \cite{SADL2021}. SADL provides both floating-point-based and integer-based inference capabilities. 
The inference of all neural networks in NNVC is based on the SADL. Table \ref{tab:sadl} summarizes the framework characteristics.

\begin{table}
\renewcommand{\arraystretch}{0.9}
\caption{Characteristics of SADL}
\label{tab:sadl}
\center
\tabcolsep0.06in
\begin{tabular}{l|l}
Language                        & Pure C++, header-only \\ 
\hline
\multirow{2}{*}{Footprint}      & $\sim$6000 lines of code, library \\
                                & $\sim$300kB, no dependency \\ 
\hline
\multirow{2}{*}{Optimization}   & SIMD at hot spots, automatic sparse  \\
                                & vector-matrix multiplication \\ 
\hline
Compatibility                   & ONNX to SADL converter \\ 
\hline
\multirow{6}{*}{Layer supports} & constants, MatMul (dense and sparse), \\ 
                                & Conv2D (strided, grouped, separated), \\
                                & Conv2DTranspose, add, mul \\
                                & MaxPool, concat, max, shape, expand, \\
                                & ReLU, PReLU, LeakyReLU, \\
                                & flatten, transpose, reshape, slicing  \\
\hline
Type support                    & float, int32, int16, int8 \\ 
\hline
Quantization                    & Support adaptive quantizer per layer \\ 
\hline
License                         & BSD 3-Clause \\ 
\hline                                       
\end{tabular}
\end{table}

\subsection{Integerized Model}
In video compression area, the fixed point implementation is crucial to allow reproducibility of the decoding, independently of the platform or environment. The SADL framework provides both floating point and fixed point implementation for all layers.

To lower the complexity of integer arithmetic of quantized model, the quantization operations required for computational layers are minimized and performed using only bit-shifting, without zero-point shifting, compared to existing method in Tensorflow \cite{jacob2017quantization} or PyTorch.

Both weights and latents tensors use the internal integer representation, e.g. int16. For  intermediate computation, the integer with twice the number of bits is used. For example, for int16 format, int32 is used for computation.
Compared to the float version, the operation are adapted as
\begin{itemize}
\item BiasAdd: $y = C \left( \left( x_0 \gg \left( q_0-q_1 \right) \right) + x_1 \right), q = q_1$
\item Add: $y = C \left( \left( x_0 \gg \left( q_0-q \right) \right) + \left( x_1 \gg \left( q_1-q \right) \right) \right), q = \min \left( q_0, q_1 \right)$
\item Mul/MatMul/Conv2D: $y = C( \sum x_0 x_1 \gg (q_1+q_i)), q=q_0-q_i$
\item Concat: $y = x_0 \gg (q_0-q) \; | \; x_1 \gg (q_1-q) \; | \; ..., q=\min(q_k)$ 
\item LeakyReLU: si $x<0, y = (\alpha x_0 ) \gg q_{\alpha}$, assuming $|\alpha| < 1$ (no overflow possible), $q=q_0$. The quantizer $q_{\alpha}$ of the slope $\alpha$ of LeakyReLU always takes  the maximum possible value to represent $\alpha$ without overflow.
\item Maximum: $y = \max(x_0, x_1 \ll (q_0-q_1)), q=q_0$
\item for layer with only one input, the output will take the quantizer of this input.
\end{itemize}
Where:
\begin{itemize}
\item $C(.)$ represents the clipping operation associated with the internal bitdepth of the latent. For example for int16 integer $C \left( x \right) =\max \left( -2^{15} + 1, \min \left( 2^{15}-1,x \right) \right)$
\item $x_0$, $x_1$: inputs
\item $y$: output
\item $q_0$ and $q_1$: shift of the quantizers. The floating value associated to a quantized input $x$ of quantizer $q$ can be recovered via $f \left( x, q \right) = x/ \left( 1 \ll q \right)$
\item $q_i$: internal shift for some layers.
\end{itemize}

\begin{figure}
\centering
\includegraphics[width=0.8\linewidth]{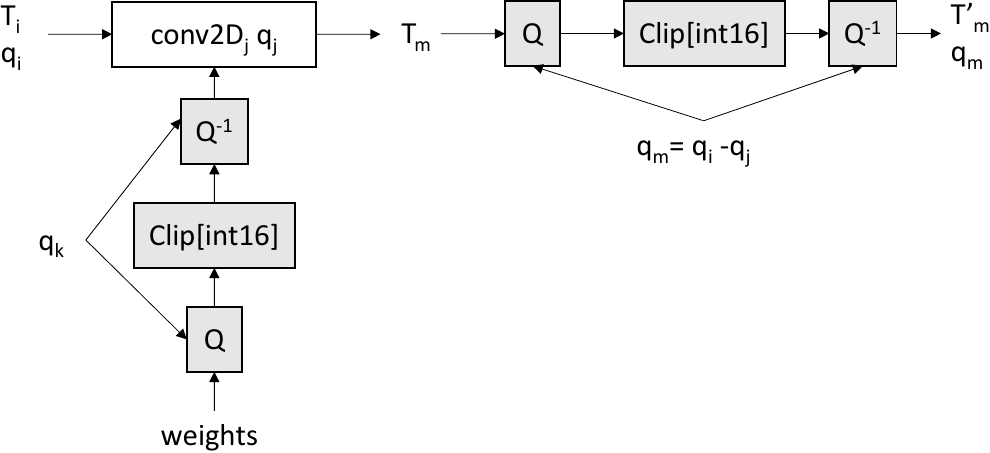}
\caption{Quantization aware convolution layer using fixed point operations.}
\label{fig_sadl_add}
\end{figure}

Conversion from trained models using floating point to integerized model can be done either using static quantization where optimal quantizers for each layer is chosen and floating point weights are converted to fixed points, or a quantization aware model training is performed. Fig. \ref{fig_sadl_add} shows an example of adaptation of the convolution layer:
quantization/clipping/dequantization stages are added for both the weights and the output of the layer.  

\subsection{Sparse Matrix Multiplication}
In order to lower the complexity of the dense layer, counted in MAC  (Multiply-Accumulate), a sparse matrix representation associated with a sparse matrix/vector is available. The sparse matrix uses a variant of the CSR (Compressed Sparse Row) representation \cite{yale}, where run-length of non-zero values is aligned on the required SIMD alignment for efficient multiply-and-add implementation on CPU: all run-length are constrained to be of a lenght modulo of 8 or 16, then each runlength is indexed in a vector and mutliply by the corresponding values of the input vector.

For example, the NN-based intra prediction model \ref{sec:intra_framework} using the sparse dense layer allows to decrease the complexity as depicted in Table \ref{table_sparse}.

\begin{table}
\renewcommand{\arraystretch}{0.9}
\caption{NN intra Sparse matrix complexity reduction}
\label{table_sparse}
\center
\tabcolsep0.06in
\begin{tabular}{|l|l|l|l|}
\hline
 \multirow{2}{*} {Model} & Dense & Sparse  & Density \\
   &  (MAC/pix) &  (MAC/pix) & \\
\hline
4 $\times$ 4 & 108300 & 7773 & 7.2\% \\
8 $\times$ 8 & 33155 & 2624 & 7.9\% \\
16 $\times$ 16 & 15627 & 1411 & 9.0\% \\
\hline

\end{tabular}
\end{table}
\subsection{Other Implementability Aspects}
In order to evaluate and compare solutions explored in JVET, complexity measurement of a model is also provided in the library. It allows to extract the real number of MACs and other operations during the inference of a particular model, independently of an underlying implementation.

A requirement of JVET evaluation is also to compare practical implementation using a reproducible environment in order to compare explored solutions to a given anchor. For these reasons, pure CPU (as opposed to GPU based) and single-threaded implementation is provided by the framework.

\section{Experimental Results}
\label{sec:experiment}
This section verifies performances of the proposed techniques in NNVC using NNVC-4.0\footnote{\url{https://vcgit.hhi.fraunhofer.de/jvet-ahg-nnvc/VVCSoftware_VTM/}}, the reference software of NNVC.
Techniques are tested under all-intra, random-access, and low-delay configurations using QP \{22, 27, 32, 37, 42\} suggested by NNVC common test conditions (CTC) \cite{Alshina2023nnvc}.
BD-rate \cite{bjontegaard2001calcuation} is adopted to measure the compression efficiency, where the quality metric bases on PSNR.
Test sequences are known as classes A1, A2, B, C, D, E, F \cite{Alshina2023nnvc}.

\begin{table*}[!t]
\renewcommand{\arraystretch}{0.9}
\caption{Overall Performance of the Proposed Techniques in NNVC-4.0 Over VTM-11.0\_{nnvc}}
\label{table_overall_results}
\centering
\tabcolsep0.06in
\begin{tabular}{c|l|ccc|ccc|ccc}
\hline
\multirow{2}{*}{Class}    &\multirow{2}{*}{Sequence} &\multicolumn{3}{c|}{Random Access} &\multicolumn{3}{c|}{Low Delay B}  &\multicolumn{3}{c}{All Intra}     \\
          \cline{3-11}
                          &                  &Y         &U          &V              &Y  &Cb &Cr                     &Y          &Cb           &Cr    \\
\hline
\multirow{3}{*}{Class A1} & Tango2           & -13.76\% & -24.82\% & -25.13\%      & - & - & -                      & -12.31\% & -30.60\% & -28.79\% \\ 
                        ~ & FoodMarket4      & -10.76\% & -16.97\% & -19.10\%      & - & - & -                      & -11.85\% & -17.89\% & -20.39\% \\ 
                        ~ & Campfire         & -9.42\%  & -11.15\% & -22.12\%      & - & - & -                      & -6.76\%  & -11.09\% & -14.86\% \\ 
\hline
\multirow{3}{*}{Class A2} & CatRobot         & -13.83\% & -25.39\% & -22.26\%       & - & - & -                      & -11.08\% & -24.68\% & -24.60\% \\ 
						~ & DaylightRoad2    & -14.84\%	&-27.13\%  &-17.57\%       & - & - & -                      & -8.75\% & -30.60\% & -18.50\% \\ 
						~ & ParkRunning3     & -7.71\%	&-10.18\%  &-10.71\%       & - & - & -                      & -8.43\% & -8.64\% & -9.05\% \\ 
\hline
\multirow{5}{*}{ClassB}   & MarketPlace      & -8.91\%  & -24.83\% & -23.98\%      & -6.24\%	&-24.54\% &-22.54\%     & -8.23\%  & -20.32\% & -22.09\% \\ 
						~ & RitualDance      & -11.42\% & -18.05\% & -26.14\%      & -8.33\%	&-14.37\% &-21.82\%     & -12.20\% & -19.13\% & -26.20\% \\ 
						~ & Cactus           & -12.27\% & -20.39\% & -19.22\%      & -8.46\%	&-18.84\% &-16.89\%     & -10.12\% & -16.47\% & -21.35\% \\ 
						~ & BasketballDrive  & -12.71\% & -27.02\% & -27.19\%      & -10.59\%   &-20.81\% &-25.19\%     & -10.74\% & -26.94\% & -28.42\% \\ 
						~ & BQTerrace        & -11.98\% & -24.86\% & -23.02\%      & -7.60\%	&-20.11\% &-16.78\%     & -6.80\%  & -22.80\% & -23.15\% \\ 
\hline
\multirow{4}{*}{classC}   & BasketballDrill  & -14.32\% & -26.08\% & -25.65\%       & -11.34\% & -17.05\% & -15.41\% & -14.29\% & -28.57\% & -29.54\% \\ 
						~ & BQMall           & -13.64\% & -26.40\% & -28.37\%       & -10.75\% & -23.81\% & -25.32\% & -11.62\% & -23.51\% & -27.56\% \\ 
						~ & PartyScene       & -13.79\% & -21.91\% & -20.65\%       & -9.94\% & -24.16\% & -20.24\%  & -7.67\%  & -15.64\% & -15.72\% \\ 
						~ & RaceHorsesC      & -9.73\% & -22.71\% & -27.81\%        & -8.42\% & -21.93\% & -25.87\%  & -7.98\%  & -19.30\% & -25.92\% \\ 
\hline
\multirow{3}{*}{classE}   & FourPeople       & - & - & -                            & -9.07\% & -16.15\% & -18.33\%  & -14.39\% & -21.57\% & -24.20\% \\ 
						~ & Johnny           & - & - & -                            & -9.29\% & -18.04\% & -22.47\%  & -14.46\% & -26.31\% & -28.51\% \\ 
						~ & KristenAndSara   & - & - & -                            & -10.12\% & -17.27\% & -20.18\% & -13.58\% & -23.98\% & -25.43\% \\ 
\hline
\hline
\multicolumn{2}{c|}{Average A1} & -11.31\% & -17.65\% & -22.11\%    & - & - & -                      & -10.31\% & -19.86\% & -21.35\% \\ 
\hline
\multicolumn{2}{c|}{Average A2} & -12.13\% &-20.90\%  &-16.85\%     & - & - & -                      & -9.42\% & -21.31\% & -17.38\% \\ 
\hline
\multicolumn{2}{c|}{Average B}  & -11.46\% & -23.03\% & -23.91\%    &-8.24\%   &-19.74\% &-20.64\% & -9.62\% & -21.13\% & -24.24\% \\ 
\hline
\multicolumn{2}{c|}{Average C}  & -12.87\% & -24.28\% & -25.62\%    & -10.11\% & -21.74\% & -21.71\% & -10.39\% & -21.75\% & -24.68\% \\ 
\hline
\multicolumn{2}{c|}{Average E}  & - & - & -                         & -9.49\% & -17.15\%  & -20.33\% & -14.14\% & -23.95\% & -26.05\% \\ 
\hline
\hline
\multicolumn{2}{c|}{\textbf{Average All}}   & \textbf{-11.94\%} & \textbf{-21.86\%} & \textbf{-22.59\%} & \textbf{-9.18\%} & \textbf{-19.76\%} & \textbf{-20.92\%} & \textbf{-10.63\%} & \textbf{-21.56\%} & \textbf{-23.02\%}  \\
\hline
\hline
\multirow{4}{*}{ClassD}   & BasketballPass         & -13.70\% & -29.29\% & -31.39\%   & -11.19\% & -25.85\% & -30.28\%    & -11.94\% & -25.46\% & -27.97\% \\ 
						~ & BQSQuare & -20.72\%    & -21.10\% & -32.78\% & -16.35\%   & -21.93\% & -29.40\% & -8.89\%     & -15.26\% & -25.31\% \\ 
						~ & BlowingBubbles         & -12.35\% & -20.12\% & -18.61\%   & -8.61\% & -24.77\% & -22.10\%     & -8.82\% & -17.43\% & -18.05\% \\ 
						~ & RaceHorses             & -12.26\% & -29.48\% & -30.42\%   & -10.69\% & -27.65\% & -28.40\%    & -10.73\% & -27.68\% & -30.22\% \\ 
\hline
\multirow{4}{*}{class F}  & BasketballDrillText    & -13.03\% & -19.92\% & -18.71\%   & -10.54\% & -16.31\% & -13.58\%    & -12.94\% & -22.71\% & -22.73\% \\ 
						~ & ArenaOfValor           & -9.72\% & -18.54\% & -18.67\%    & -6.10\%	 &-16.77\%	&-12.47\%    & -9.53\% & -20.41\% & -21.16\% \\ 
						~ & SlideEditting          & 0.34\% & -1.84\% & -2.32\%       & -0.66\% & -4.38\% & -6.24\%       & 0.04\% & -4.07\% & -5.17\% \\ 
						~ & SlideShow              & -4.91\% & -19.67\% & -13.83\%    & -7.19\% & -18.95\% & -10.97\%     & -6.80\% & -21.20\% & -14.72\% \\ 
\hline
\hline
\multicolumn{2}{c|}{Average D}  & -14.76\% & -25.00\% & -28.30\%    & -11.71\% & -25.05\% & -27.55\%     & -10.10\% & -21.46\% & -25.39\% \\ 
\hline
\multicolumn{2}{c|}{Average F}  & -6.83\% & -14.99\% & -13.38\%     & -6.12\%  &-14.10\%  &-10.81\%     & -7.31\% & -17.10\% & -15.94\% \\ 
\hline
\end{tabular}
\end{table*}

\begin{figure*}
\centering
\subfigure[ ]
{
\includegraphics[width=0.25\linewidth]{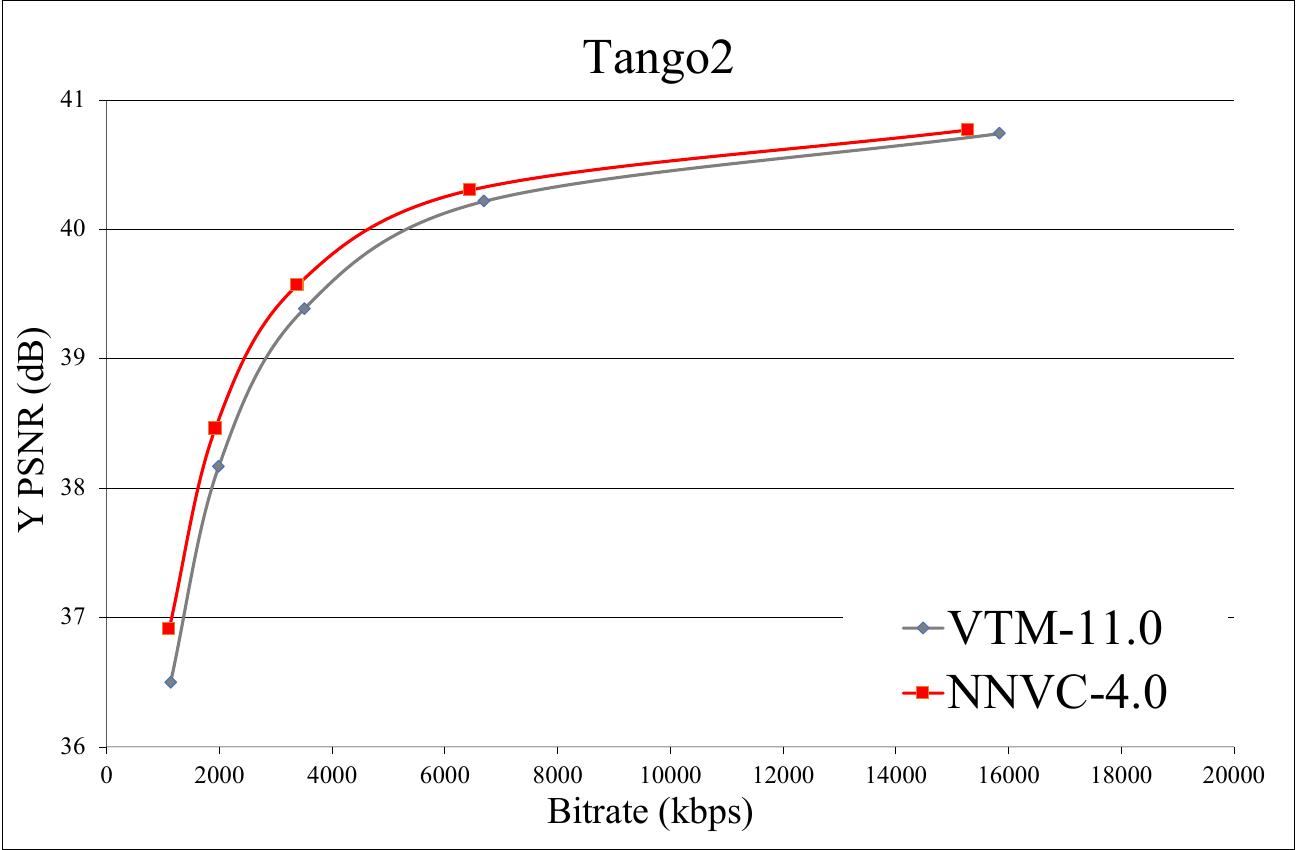}
}
\subfigure[ ]
{
\includegraphics[width=0.25\linewidth]{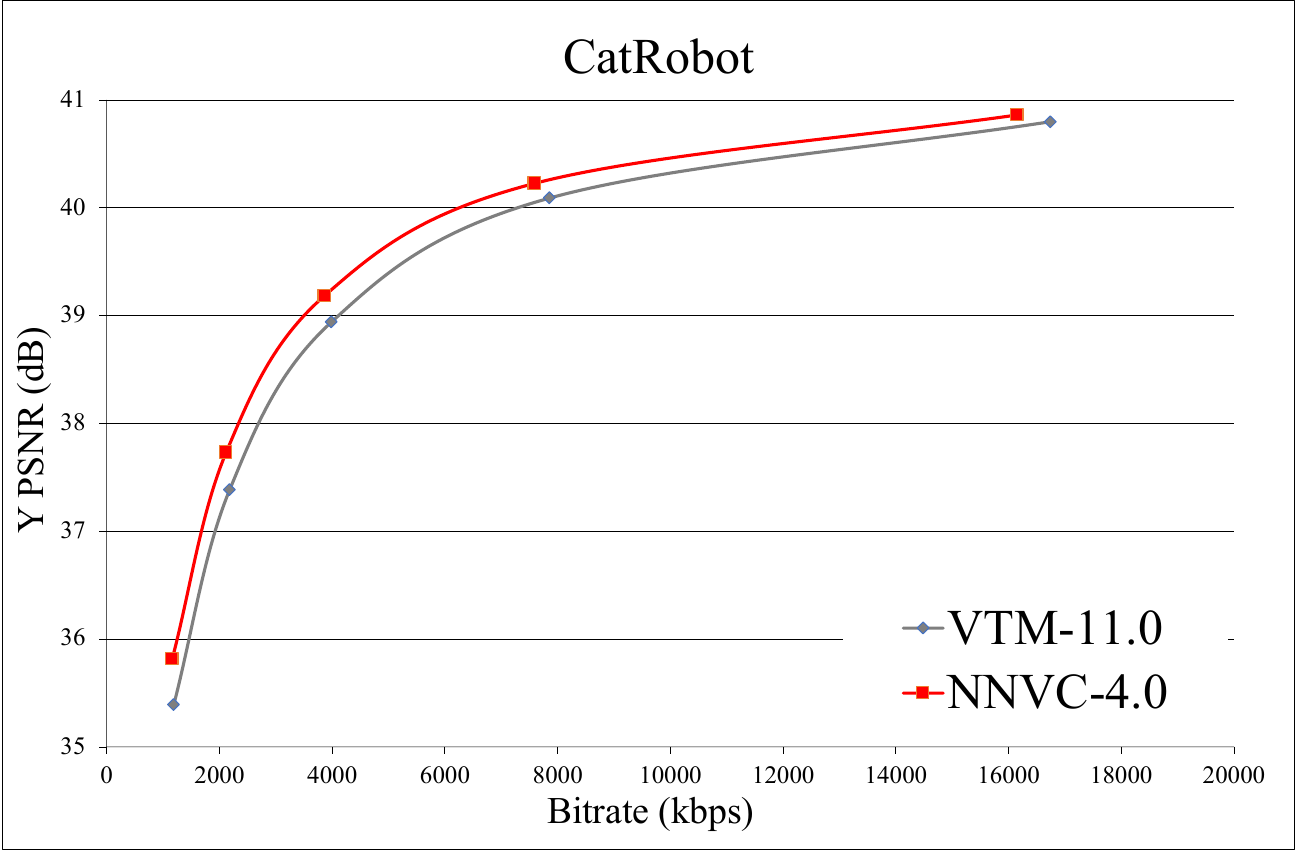}
}
\subfigure[ ]
{
\includegraphics[width=0.25\linewidth]{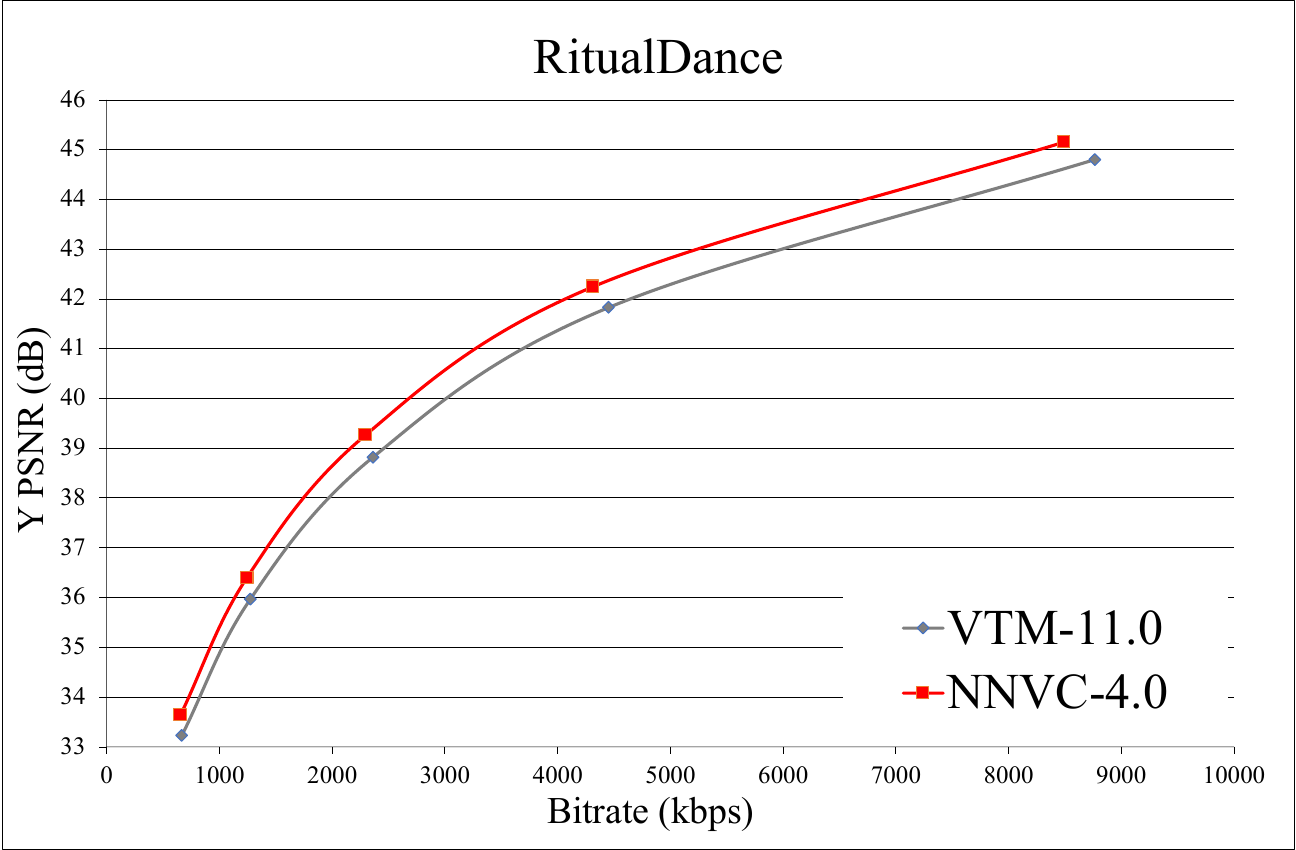}
}
\subfigure[ ]
{
\includegraphics[width=0.25\linewidth]{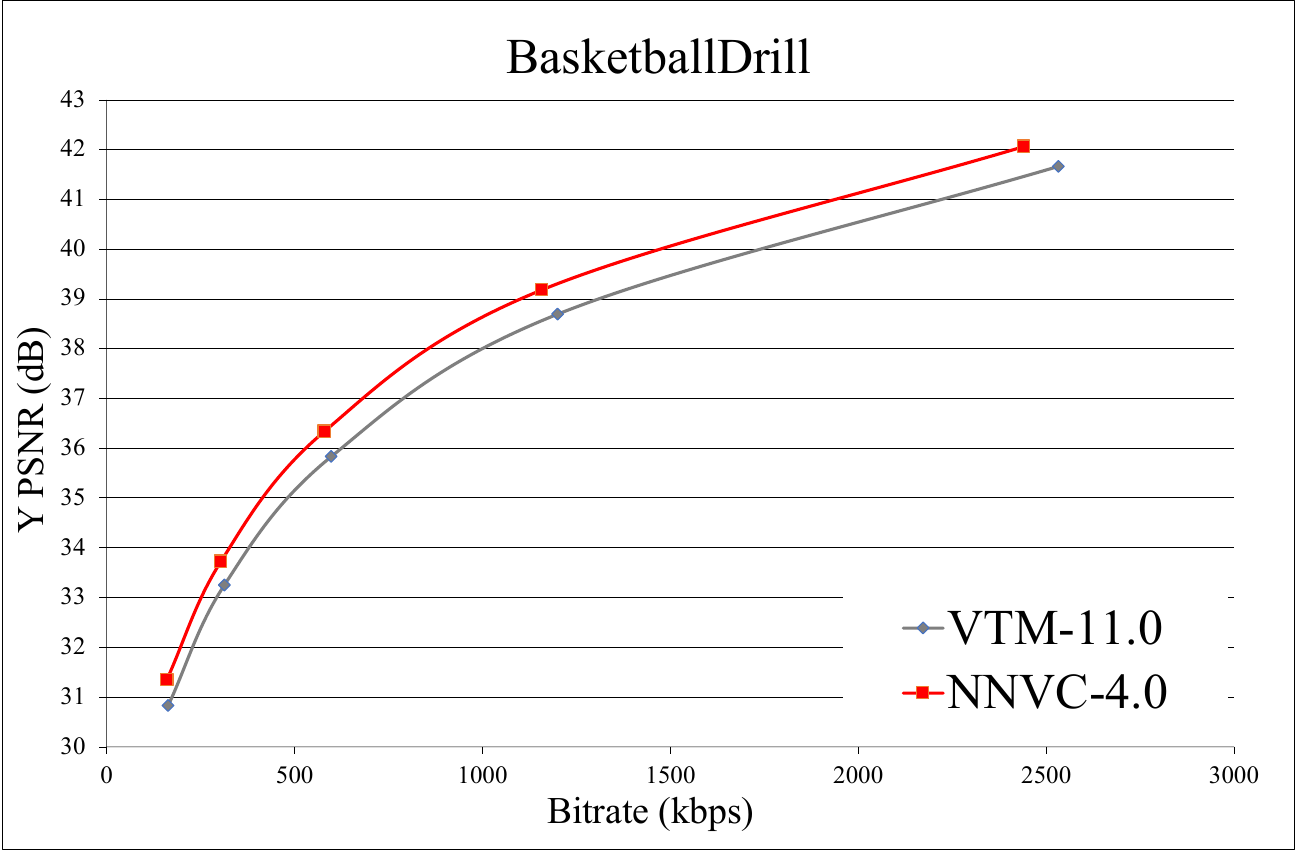}
}
\subfigure[ ]
{
\includegraphics[width=0.25\linewidth]{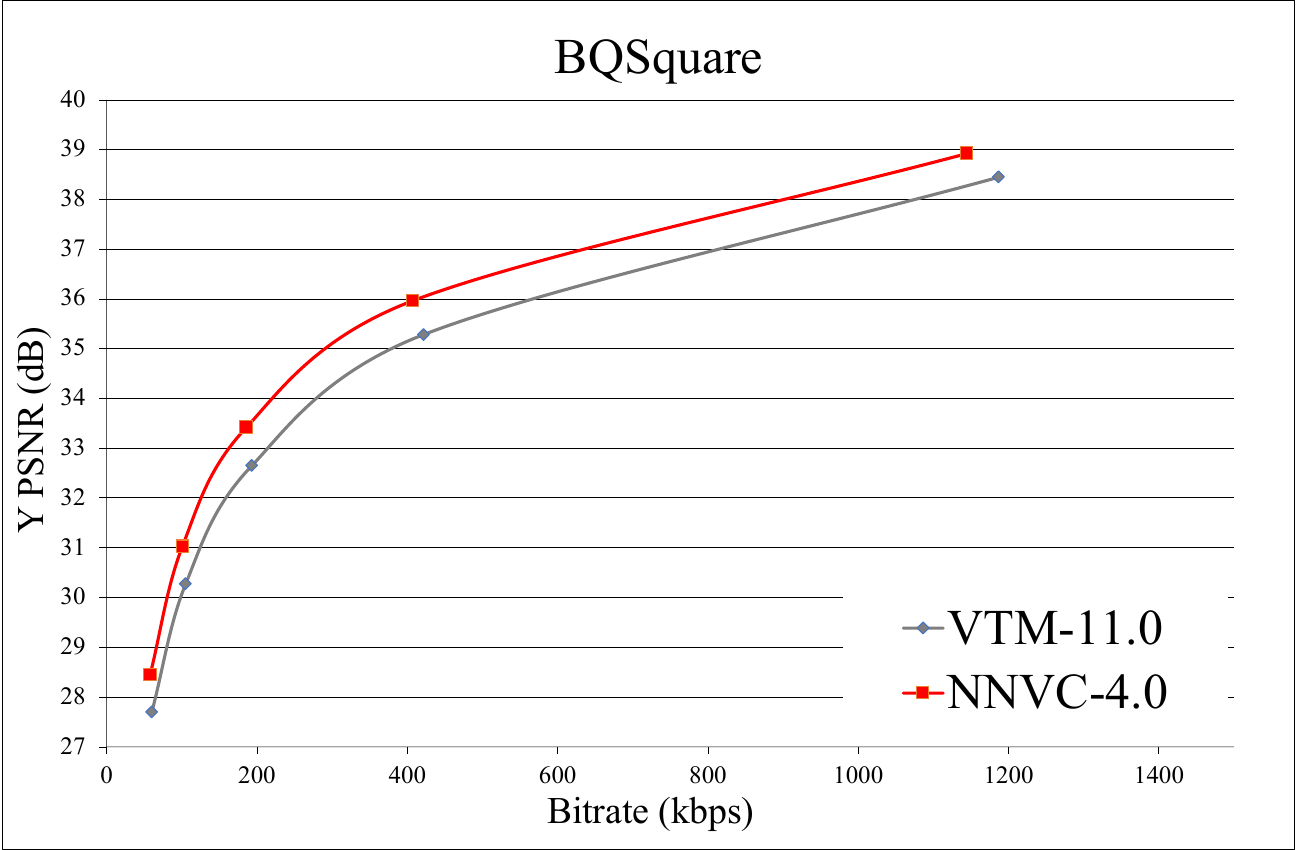}
}
\subfigure[ ]
{
\includegraphics[width=0.25\linewidth]{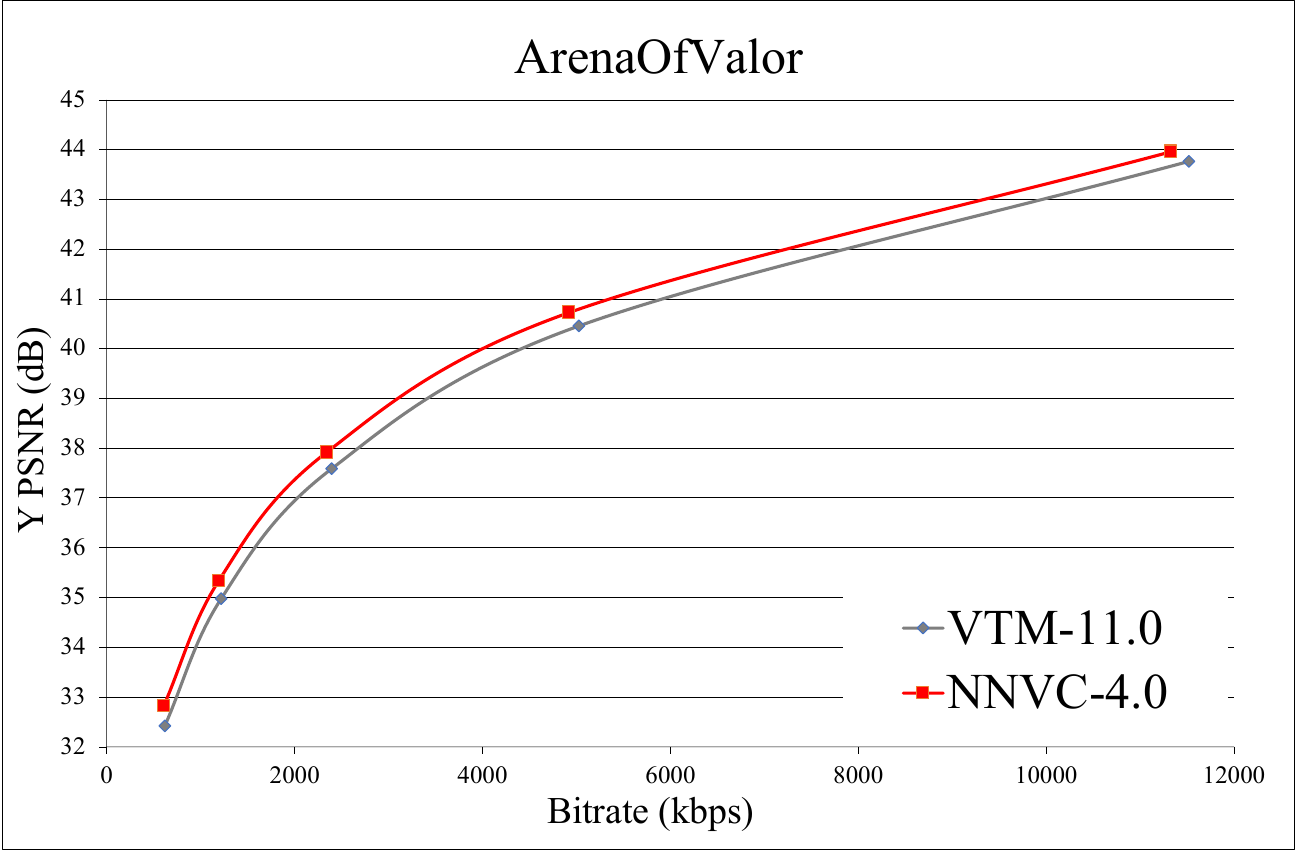}
}
\caption{Rate-distortion (R-D) curves of several sequences in different classes: (a) Tango2 (Class A1), (b) CatRobot (Class A2), (c) RitualDance (Class B), (d) BasketballDrill (class C), (e) BQSquare (Class D), and (f) ArenaOfValor (Class F).}
\label{fig_rd_curves}
\end{figure*}

\begin{table}[!t]
\renewcommand{\arraystretch}{0.9}
\caption{Computational Time of the Proposed Techniques in NNVC-4.0 relative to VTM-11.0\_{nnvc}.}
\label{table_complexity}
\centering
\tabcolsep0.06in

\begin{tabular}{c|cc|cc|cc}
\hline
\multirow{2}{*}{Class}  &\multicolumn{2}{c|}{Random Access} &\multicolumn{2}{c|}{Low Delay B}  &\multicolumn{2}{c}{All Intra}      \\
       \cline{2-7}
                        &Enc.        &Dec.             &Enc.        &Dec.                    &Enc.        &Dec.  \\  
\hline
Class A1                & 2.1 & 351.4 & -  & -          & 2.9  & 260.4  \\ 
\hline
Class A2                & 2.1 & 333.7 & -  & -          & 2.5  & 206.9  \\ 
\hline
Class B                 & 2.1  & 328.0  & 2.1 & 357.0   & 2.5  & 200.0  \\ 
\hline
Class C                 & 2.0  & 296.6  & 1.8 & 304.8   & 2.3  & 135.0  \\ 
\hline
Class E                 & -  & -        & 3.1 & 242.3   & 2.5  & 225.6  \\ 
\hline
\hline
Average                 & 2.1 & 324.9 & 2.2 & 307.4 & 2.5  & 196.5  \\ 
\hline
\hline
Class D                 & 2.0  & 245.0  & 1.7  & 244.6  & 2.3  & 122.1  \\ 
\hline
Class F                 & 2.7  & 133.5  & 2.5  & 162.9  & 1.8  & 162.2 \\ 
\hline

\end{tabular}
\end{table}

\subsection{Overall Results}
\label{sec_overall_results}
Table \ref{table_overall_results} gives overall performances of the proposed techniques in NNVC-4.0 over VTM-11.0\_{nnvc}.
Following JVET common test conditions \cite{bossen2018jvet}, we exclude classes D and F when computing the overall average.
As can be observed, NNVC-4.0 with the proposed techniques outperforms VTM-11.0\_{nnvc} significantly, achieving on average \{11.94\%, 21.86\%, 22.59\%\}, \{9.18\%, 19.76\%, 20.92\%\}, and \{10.63\%, 21.56\%, 23.02\%\} BD-rate reductions for \{Y, Cb, Cr\}, under random-access, low-delay, and all-intra configurations, respectively.
The proposed NN models are tuned on natural contents, therefore gains are limited on Class F containing screen content sequences.

Fig. \ref{fig_rd_curves} shows R-D curves for sequences from different classes.
Trends can be observed that NNVC-4.0 offers relatively higher coding gains at middle bit-rates.
The phenomenon may be related to the distortion characteristics at different bit-rates.
The low bit-rate tends to yield larger distortion, making it more difficult to infer the lost details from existing contexts while the high bit-rate usually means low distortion level, leaving limited space for further reduction. 

Currently, the implementation of the NN-based models is not fully optimized and is CPU-based, thus the encoding/decoding time of NNVC-4.0 is much longer than that of the highly optimized VVC reference software.
Table \ref{table_complexity} presents the computational time comparison. 
Encoding complexities are 2.1, 2.2, and 2.5 times for random-access, low-delay, and all-intra cases, respectively.
Regarding decoding complexities, they are 324.9, 307.4, and 196.5 times for random-access, low-delay, and all-intra cases, respectively.
Note that in real applications, inference of NN-based models could be accelerated significantly with more efficient architectures such as GPUs (graphics processing units), TPUs (tensor processing units, a kind of application-specific integrated circuits) or full custom ASIC silicon.
Besides running time, number of total parameters and multiply-accumulates (MACs) are important measurements concerning complexity considered in NNVC.
Details regarding these measurements could be found in Table \ref{tab:intra_inference} and Table \ref{tab:nnlf_inference}.

\begin{table*}[!t]
\renewcommand{\arraystretch}{0.9}
\caption{Performances of NNVC-4.0 configured in different modes (for luma component only)}
\label{table_ablation_results}
\centering
\tabcolsep0.06in
\begin{tabular}{c|l|cc|cc|cc|cc|cc}
\hline
\multirow{2}{*}{Class}    &\multirow{2}{*}{Sequence} &\multicolumn{2}{c|}{$M_1$} &\multicolumn{2}{c|}{$M_2$} &\multicolumn{2}{c}{$M_3$} &\multicolumn{2}{c|}{$M_4$} &\multicolumn{2}{c}{$M_5$}     \\
          \cline{3-12}
                          &                 &RA       &AI         &RA        &AI        &RA        &AI         &RA       &AI          &RA        &AI     \\
\hline
\multirow{3}{*}{Class A1} & Tango2          & -2.54\% & -5.47\%   & -10.78\% & -7.52\%  & -12.72\% & -12.20\%  & -13.65\% & -12.20\%  & -13.76\% & -12.31\% \\ 
                        ~ & FoodMarket4     & -2.18\% & -4.80\%   & -8.28\%  & -7.82\%  & -9.98\%  & -11.72\%  & -10.67\% & -11.72\%  & -10.76\% & -11.85\% \\ 
                        ~ & Campfire        & -2.24\% & -2.66\%   & -7.13\%  & -4.30\%  & -8.90\%  & -6.65\%   & -9.25\%  & -6.65\%   & -9.42\%  & -6.76\% \\ 
\hline
\multirow{3}{*}{Class A2} & CatRobot        & -1.95\% & -3.62\%   & -11.01\% & -7.73\%  & -12.51\% & -10.82\%  & -13.47\% & -10.82\%  & -13.83\% & -11.08\% \\ 
                        ~ & DaylightRoad2   & -1.45\% & -2.79\%   & -12.43\% & -6.04\%  & -13.52\% & -8.47\%   & -14.41\% & -8.47\%   & -14.84\% & -8.75\% \\ 
                        ~ & ParkRunning3    & -0.93   & -2.05\%   & -6.24\% & -6.34\%   & -7.01\%    & -8.23\%   & -7.51\% & -8.23\%      & -7.71\% & -8.43\% \\ 
\hline
\multirow{3}{*}{Class B}  & MarketPlace     & -1.10\% & -2.23\%   & -7.10\% & -6.05\%   & -7.91\%  & -7.96\%   & -8.60\% & -7.96\%    & -8.91\%  & -8.23\% \\ 
						~ & RitualDance     & -2.19\% & -4.03\%   & -8.88\% & -8.81\%   & -10.59\% & -11.94\%  & -11.21\% & -11.94\%  & -11.42\% & -12.20\% \\ 
						~ & Cactus          & -2.14\% & -3.37\%   & -9.21\% & -7.07\%   & -10.85\% & -9.89\%   & -11.91\% & -9.89\%   & -12.27\% & -10.12\% \\ 
						~ & BasketballDrive & -2.14\% & -4.17\%   & -10.71\% & -7.10\%  & -12.46\% & -10.61\%  & -12.54\% & -10.61\%  & -12.71\% & -10.74\% \\ 
						~ & BQTerrace       & -1.52\% & -2.38\%   & -9.60\% & -4.44\%   & -10.91\% & -6.60\%   & -11.32\% & -6.60\%   & -11.98\% & -6.80\% \\ 
\hline
\multirow{3}{*}{Class C}  & BasketballDrill & -2.19\% & -4.10\%   & -11.23\% & -10.54\% & -12.89\% & -13.85\%  & -13.72\% & -13.85\%  & -14.32\% & -14.29\% \\ 
						~ & BQMall          & -1.60\% & -3.29\%   & -11.41\% & -8.66\%  & -12.71\% & -11.34\%  & -13.34\% & -11.34\%  & -13.64\% & -11.62\% \\ 
						~ & PartyScene      & -1.50\% & -2.51\%   & -11.73\% & -5.22\%  & -12.87\% & -7.41\%   & -13.36\% & -7.41\%   & -13.79\% & -7.67\% \\ 
						~ & RaceHorsesC     & -1.47\% & -2.88\%   & -8.28\% & -5.31\%   & -9.46\%  & -7.79\%   & -9.53\%  & -7.79\%   & -9.73\%  & -7.98\% \\ 
\hline
\multirow{3}{*}{Class E}  & FourPeople      & - & -4.74\%         & - & -10.42\%        & -        & -14.08\%  & -        & -14.08\%  & -        & -14.39\% \\ 
						~ & Johnny          & - & -5.21\%         & - & -10.12\%        & -        & -14.26\%  & -        & -14.26\%  & -        & -14.46\% \\ 
						~ & KristenAndSara  & - & -4.72\%         & - & -9.50\%         & -        & -13.30\%  & -        & -13.30\%  & -        & -13.58\% \\ 
\hline
\hline
\multicolumn{2}{c|}{Average A1}  & -2.32\% & -4.31\%              & -8.73\%  & -6.55\%  & -10.53\% & -10.19\%  & -11.19\% & -10.19\%  & -11.31\% & -10.31\% \\ 
\hline
\multicolumn{2}{c|}{Average A2}  & -1.44\% & -2.82\%              & -9.89\%  & -6.70\%  & -11.01\%   & -9.17\%   & -11.80\% & -9.17\%     & -12.13\% & -9.42\% \\ 
\hline
\multicolumn{2}{c|}{Average B}   & -1.82\% & -3.24\%              & -9.10\%  & -6.69\%  & -10.54\% & -9.40\%   & -11.12\% & -9.40\%   & -11.46\% & -9.62\% \\ 
\hline
\multicolumn{2}{c|}{Average C}   & -1.69\% & -3.20\%              & -10.66\% & -7.43\%  & -11.98\% & -10.10\%  & -12.49\% & -10.10\%  & -12.87\% & -10.39\% \\ 
\hline
\multicolumn{2}{c|}{Average E}   & -       & -4.89\%              & -        & -10.01\% & -        & -13.88\%  & -        & -13.88\%  & -        & -14.14\% \\ 
\hline
\hline
\multicolumn{2}{c|}{\textbf{Average All}} & \textbf{-1.81\%} & \textbf{-3.61\%} & \textbf{-9.60\%} & \textbf{-7.39\%} & \textbf{-11.02\%} & \textbf{-10.40\%} & \textbf{-11.63\%} & \textbf{-10.40\%} & \textbf{-11.94\%} & \textbf{-10.63\%} \\ 
\hline
\hline
\multirow{3}{*}{Class D}  & BasketballPass      & -1.52\% & -3.27\% & -11.57\% & -8.98\% & -12.74\% & -11.69\% & -13.38\% & -11.69\% & -13.70\% & -11.94\% \\ 
						~ & BQSQuare            & -1.08\% & -2.25\% & -19.05\% & -6.61\% & -20.06\% & -8.57\% & -20.15\% & -8.57\% & -20.72\% & -8.89\% \\ 
						~ & BlowingBubbles      & -1.40\% & -2.94\% & -10.13\% & -6.07\% & -11.26\% & -8.54\% & -12.05\% & -8.54\% & -12.35\% & -8.82\% \\ 
						~ & RaceHorses          & -1.69\% & -3.52\% & -10.24\% & -7.56\% & -11.49\% & -10.46\% & -11.84\% & -10.46\% & -12.26\% & -10.73\% \\ 
\hline
\multirow{3}{*}{Class F}  & BasketballDrillText & -1.78\% & -3.58\% & -10.70\% & -9.64\% & -12.03\% & -12.55\% & -12.49\% & -12.55\% & -13.03\% & -12.94\% \\ 
						~ & ArenaOfValor        & -1.55\% & -3.12\% & -7.46\% & -6.58\% & -8.74\% & -9.28\% & -9.35\% & -9.28\% & -9.72\% & -9.53\% \\ 
						~ & SlideEditting       & -0.31\% & -0.23\% & 0.74\% & 0.44\% & 0.39\% & 0.09\% & 0.37\% & 0.09\% & 0.34\% & 0.04\% \\ 
						~ & SlideShow           & -0.79\% & -1.89\% & -4.52\% & -4.93\% & -5.25\% & -6.80\% & -4.74\% & -6.80\% & -4.91\% & -6.80\% \\ 
\hline
\hline
\multicolumn{2}{c|}{Average D} & -1.42\% & -3.00\% & -12.75\% & -7.30\% & -13.89\% & -9.81\% & -14.35\% & -9.81\% & -14.76\% & -10.10\% \\ 
\hline
\multicolumn{2}{c|}{Average F} & -1.11\% & -2.21\% & -5.49\% & -5.17\% & -6.41\% & -7.14\% & -6.55\% & -7.14\% & -6.83\% & -7.31\% \\ 
\hline
\end{tabular}
\end{table*}

\subsection{Ablation Test}
\label{sec_ablation_results}
Table \ref{table_ablation_results} gives performances of NNVC-4.0 configured in different modes.
BD-rate changes are shown for luma component in all-intra and random-access settings.
NN tools enabled in each mode are explained below,
\begin{itemize}
 \item $M_1$, NN-based intra prediction.
 \item $M_2$, basic NN-based in loop filter, i.e. without temporal filtering (Section \ref{sec:temporal_filtering}) and encoder optimization (Section \ref{sec:EncNnlfOpt}).
 \item $M_3$, NN-based intra prediction, basic NN-based in loop filter.
 \item $M_4$, NN-based intra prediction, basic NN-based in loop filter, temporal filtering (Section \ref{sec:temporal_filtering}).
 \item $M_5$, NN-based intra prediction, basic NN-based in loop filter, temporal filtering, encoder optimization (Section \ref{sec:EncNnlfOpt}).
\end{itemize}

Compared with VTM-11.0\_{nnvc}, NN-based intra prediction and basic in-loop filter provide on average \{1.81\%, 3.61\%\} and \{9.60\%, 7.39\%\} BD-rate reductions for the luma component under random-access and all-intra settings respectively (as observed from columns $M_1$ and $M_2$).
By combing the two tools, BD-rate reductions rise to \{11.02\%, 10.40\%\} for random-access and all-intra settings as shown in column $M_3$.
Comparison of columns $M_1$, $M_2$, and $M_3$ suggests gains of NN-based intra prediction and in-loop filtering are almost additive, yet the tools are trained and optimized separately.
Results in columns $M_4$ and $M_5$ can reflect additional improvements due to temporal filtering and encoder optimization techniques, i.e. an approximate BD-rate reduction of 0.6\% from temporal filter and 0.3\% from encoder optimization in the random-access setting.

\section{Conclusion}
\label{sec:conclusion}
Joint Video Experts Team of ITU-T SG 16 WP 3 and ISO/IEC JTC 1/SC29 is working together on an exploration study to evaluate potential NNVC technology beyond the capabilities of VVC.
The exploration activity has identified two promising NN-based coding tools as an enhancement of existing intra prediction and in-loop filtering techniques in VVC design. 
This paper introduced technical features, encoding methods, and training methods of some of these tools.
Implementation of these tools in NNVC reference software is based on SADL.
Effectivenesses of the NNVC techniques have been verified by the experimental results about NNVC-4.0, i.e. \{11.94\%, 21.86\%, 22.59\%\}, \{9.18\%, 19.76\%, 20.92\%\}, and \{10.63\%, 21.56\%, 23.02\%\} BD-rate reductions on average for \{Y, Cb, Cr\} compared with VVC under random-access, low-delay, and all-intra settings respectively.
Future works on complexity reduction and other competitive NN tools are encouraged.

\ifCLASSOPTIONcaptionsoff
  \newpage
\fi

\bibliographystyle{ieeetr}
\bibliography{refs}




\end{document}